\documentclass[sn-nature, iicol, pdflatex, crop]{sn-jnl}

\usepackage{graphicx}%
\usepackage{multirow}%
\usepackage{amsmath,amssymb,amsfonts}%
\usepackage{amsthm}%
\usepackage{mathrsfs}%
\usepackage[title]{appendix}%
\usepackage{xcolor}%
\usepackage{textcomp}%
\usepackage{manyfoot}%
\usepackage{booktabs}%
\usepackage{algorithm}%
\usepackage{algorithmicx}%
\usepackage{algpseudocode}%
\usepackage{listings}%

\usepackage{bm}
\usepackage{bbm}
\usepackage{upgreek}

\newcommand{\pars}{\bm{\xi}}
\newcommand{\M}{\bm{M}_s}
\renewcommand{\dim}[0]{d}

\unnumbered

\begin{document}

\title{The polyhedral structure underlying programmable self-assembly}

\author[1]{\fnm{Maximilian C.} \sur{H\"ubl}}\email{maximilian.huebl@ist.ac.at}

\author[2]{\fnm{Thomas E.} \sur{Videb{\ae}k}}\email{videbaek@brandeis.edu}

\author[2]{\fnm{Daichi} \sur{Hayakawa}}\email{dhayakawa@brandeis.edu}

\author*[2]{\fnm{W. Benjamin} \sur{Rogers}}\email{wrogers@brandeis.edu}

\author*[1]{\fnm{Carl P.} \sur{Goodrich}}\email{carl.goodrich@ist.ac.at}

\affil[1]{\orgname{Institute of Science and Technology Austria (ISTA)}, \orgaddress{\street{Am Campus 1}, \city{Klosterneuburg}, \postcode{3400}, \country{Austria}}}

\affil[2]{\orgdiv{Martin A. Fisher School of Physics}, \orgname{Brandeis University}, \orgaddress{\city{Waltham}, \postcode{02453}, \state{Massachusetts}, \country{USA}}}

\abstract{
Experiments have reached a monumental capacity for designing and synthesizing microscopic particles for self-assembly, making it possible to precisely control particle concentrations, shapes, and interactions. However, more physical insight is needed before we can take full advantage of this vast design space to assemble nanostructures with complex form and function. 
Here we show how a significant part of this design space can be quickly and comprehensively understood by identifying a class of thermodynamic constraints that act on it.
These thermodynamic constraints form a high-dimensional convex polyhedron that determines which nanostructures can be assembled at high equilibrium yield and reveals limitations that govern the coexistence of structures, which we verify through detailed, quantitative assembly experiments of nanoscale particles synthesized using DNA origami. Strong experimental agreement confirms the importance of the polyhedral structure and motivates its use as a predictive tool for the rational design of self-assembly. These results uncover fundamental physical relationships underpinning many-component programmable self-assembly in equilibrium and form the basis for robust inverse-design, applicable to a wide array of systems from biological protein complexes to synthetic nanomachines.
}


\maketitle

\begin{figure*}
\centering
\includegraphics[width=\linewidth]{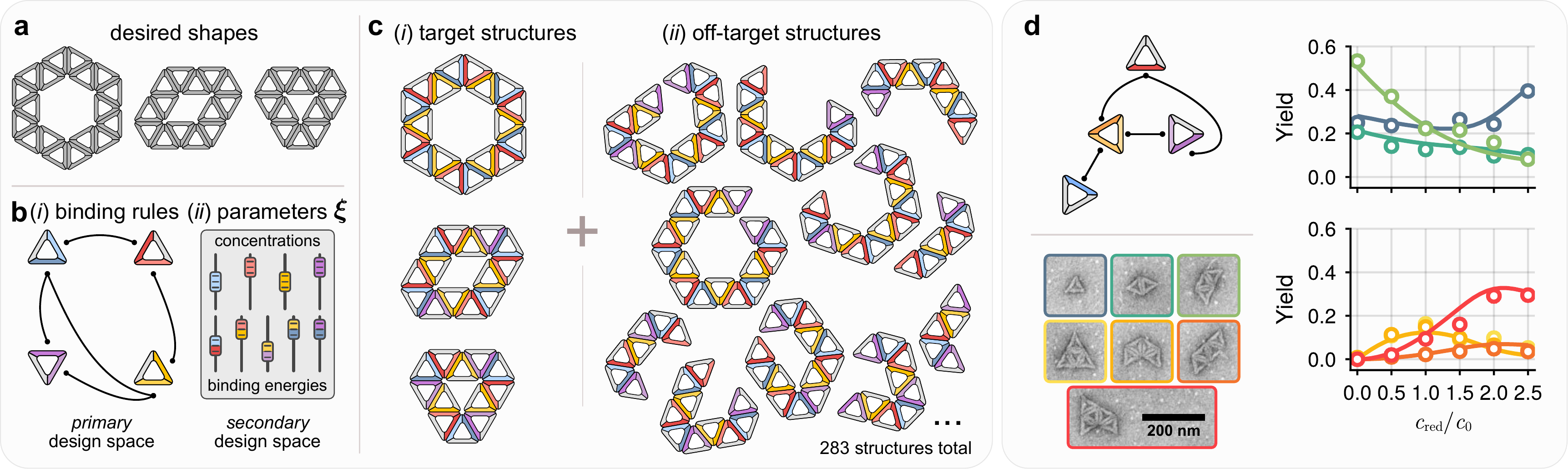}
    \caption{Design spaces of programmable self-assembly.
    (a) The design process often starts with desired target shapes, here three rings, that should be assembled.
    (b(i)) Binding rules that allow the assembly of the desired shapes from few components. Allowed bonds between triangle sides are indicated by black lines. Gray sides are inert. The space of all binding rules forms the (discrete) \emph{primary} design space. (ii) For a given set of binding rules, it is possible to change particle concentrations and binding energies, which together form the (continuous) \emph{secondary} design space, here illustrated with sliders tuning the individual parameters.
    (c) Assuming in-plane assembly with rigid binding, the binding rules shown in (b) allow the formation of the three desired target shapes (i) and 280 additional, off-target structures consisting of chimeras and incompletely assembled structures (ii).
    (d) Experimental validation of yield calculations. Measured (points) and theoretical (lines) yields of all observed structure shapes resulting from the triangular particles and binding rules shown on the top. Colors of the points and lines correspond to the color outline around the seven observed structure shapes. Structure yields are shown as a function of the concentration $c_\mathrm{red}$ of the red particle species. Each other particle species is supplied at concentration $c_0$. Data is separated into two plots for better visibility. Error bars show the standard error of the measured yield, and are generally smaller than the plot markers. See Methods and SI for details.
    }
    \label{fig:intro}
\end{figure*}

Programmable self-assembly holds enormous potential for the construction of complex nanostructures at scale. The past decade has seen the development of an array of advanced experimental platforms for designing and synthesizing particles with tunable and specific interactions that, guided by theoretical design principles, lead to the formation of precisely defined finite-size nanostructures~\cite{Mirkin.1996, Meng.2010, Wang.2012, Wang.2015, Rogers.2016, Jacobs.2024, Park.2006, Wei.2012, Ke.2012, Evans.2024, Bai.2012, Funke.2016, Hayakawa.2022pqs, Hayakawa.2024, Sigl.2021, Wei.2024, Videbaek.2025, Sacanna.2010, Huntley.2016, Huang.2016, Boyken.2016, Niu.2019}. 
However, synthetic self-assembly is still no match for self-assembly in biology, which is capable of assembling a multitude of complex structures from shared components and can steer the assembly outcome based on external cues.
In contrast, the vast majority of work in synthetic programmable assembly starts with a single, static target structure, and achieving high-yield assembly often requires the creation of a large number of distinct and individually addressable particle species~\cite{Murugan.2015, Zeravcic.20140h, Jacobs.2024, Ke.2012, Wei.2012}. 
In addition to being highly uneconomical~\cite{Hayakawa.2024, Bohlin.2023, Huebl.2024}, this severely limits our ability to assemble multiple structures simultaneously or design multifarious or reconfigurable assemblies~\cite{Murugan.2015kjd, Osat.2023, Evans.2024}.

Moving beyond the design of a single static structure requires less restrictive binding rules. For example, Fig.~\ref{fig:intro} shows three target ring-like shapes (a) and a set of allowed bonds between four particle types (b(i)) that allow for the assembly of these shapes (c(i)). However, these binding rules are also compatible with 280 other structures (c(ii)), meaning that further design parameters are necessary to achieve any reasonable level of control over the assembly outcome.
Fortunately, experimental platforms are able to do more than just control which bonds are allowed; many can independently adjust the binding energies of each bond type and introduce the particle species at different concentrations~\cite{Murugan.2015, Hayakawa.2024, Videbaek.2024}. These binding energies and particle concentrations form a secondary design space (Fig.~\ref{fig:intro}b(ii)), defined for fixed binding rules, that has not been systematically explored.

In this paper, we show how to fully and robustly understand this secondary design space. 
More specifically, we show that equilibrium statistical mechanics implies the existence of a series of thermodynamic constraints that together have the mathematical structure of a high-dimensional convex polyhedral cone. The nature of this cone dictates theoretically allowed yields for a given choice of binding rules, enabling us to quickly and exactly determine whether a desired assembly outcome is possible or not.
Furthermore, this polyhedral cone lets us identify ``necessary chimeras'' -- off-target structures that are thermodynamically unavoidable given a particular target -- and reveals low-dimensional relationships between the relative yields of coexisting structures. 

To test the practical utility of our theory, we design and synthesize a set of DNA-origami particles, and perform quantitative experiments to self-assemble the set of ring-shaped structures shown in Fig.1a. Without \emph{a priori} knowledge of the binding energies or particle concentrations, and without modeling the details of the interactions, our theory is able to quantitatively predict the possible relative yields of the coexisting structures under various conditions. Taken together, our results demonstrate an internal logic to programmable self-assembly in equilibrium that leads to far-reaching predictions about physically possible assembly outcomes, which are independent of the microscopic details of the assembling particles.
This broad generality makes our results the basis for a robust framework for economical inverse design in a wide range of experimental settings, from lock-and-key colloids~\cite{Sacanna.2010} to protein complexes~\cite{Huang.2016} and DNA nanoparticles~\cite{Jacobs.2024}.

\subsection{Equilibrium statistical mechanics predicts experimental yields}
We consider the equilibrium self-assembly of finite-size structures out of smaller programmable particles. As is the case in many experimental settings, we define a number of particle species (usually between 1-20), and specific short-ranged interactions lead to the formation of bonds only according to the binding rules (e.g. Fig.~\ref{fig:intro}b(i)). 
While significant experimental and theoretical work has focused on altering these binding rules to control the assembly process~\cite{hormoz2011, Zeravcic.20140h, Bohlin.2023, Huebl.2024, Ke.2012, Wei.2012, Hayakawa.2022pqs, Hayakawa.2024, Videbaek.2022, Videbaek.2024}, we will fix the binding rules and instead consider the impact of altering the binding energy of each bond type and the chemical potential (or equivalently the particle concentration) of each particle species (Fig.~\ref{fig:intro}b(ii)). 

To proceed, we combine all binding energies and chemical potentials into a single vector $\pars$, which we express in units of $k_\mathrm{B}T$, the Boltzmann constant times the temperature. Furthermore, let $d$ be the length of this vector, {\it i.e.} the combined number of independently adjustable binding energies and chemical potentials. Following a straightforward statistical mechanics formulation of the assembly outcome~\cite{Klein.2018, Curatolo.2023, Holmes-Cerfon.2013, Meng.2010, Israelachvili2011}, the equilibrium number density of a particular structure $s$ is given by the mass action law~\cite{Curatolo.2023, Hagan.2021, Kampen1992}
\begin{equation}\label{eq:rho}
    \rho_s(\pars) = \Omega_s \, e^{\M \cdot \pars} \,,
\end{equation}
where $\Omega_s$ is a positive pre-factor related to the symmetry and entropy of $s$ and depends on the system-specific details of the binding interactions.
$\M \in \mathbb{N}^d$ is a vector listing the number of each particle species and bond type in $s$ (see Methods).
Importantly, for particles assembling with short-ranged interactions, the parameters $\pars$ only enter \emph{linearly} in the exponential in Eq.~\eqref{eq:rho}, which has important consequences, as we will see.

We assume for now that the set of all possible structures that are allowed by the binding rules, $\mathcal{S}$, is finite and has been computed, for example through the methods in Refs.~\cite{Huebl.2024, Arkus.2011, McMullen.2022}.
By summing over all possible structures, we can then compute the equilibrium yield of every structure via 
\begin{equation}\label{eq:yields}
    Y_s(\pars) = \frac{\rho_s(\pars)}{\sum_{s' \in \mathcal{S}} \rho_{s'}(\pars)}\,.
\end{equation}
Figure~\ref{fig:intro}d and the related discussion in Methods demonstrate that Eq.~\eqref{eq:rho} and \eqref{eq:yields} lead to accurate and robust predictions of experimental yields.

\subsection{Thermodynamic constraints and their polyhedral structure}
While Eq.~\eqref{eq:rho} and Eq.~\eqref{eq:yields} enable us to predict structure yields from the design parameters, solving the inverse problem, {\it i.e.} finding parameters that maximize the yield of a desired structure(s), is highly non-trivial.
In fact, it is often unclear \emph{if} high-yield assembly is possible at all.
To understand when, why, and how high-yield assembly is possible, we need to dive deeper into the mathematical structure implied by Eq.~\eqref{eq:rho}.

\begin{figure*}
    \centering
    \includegraphics[width=\textwidth]{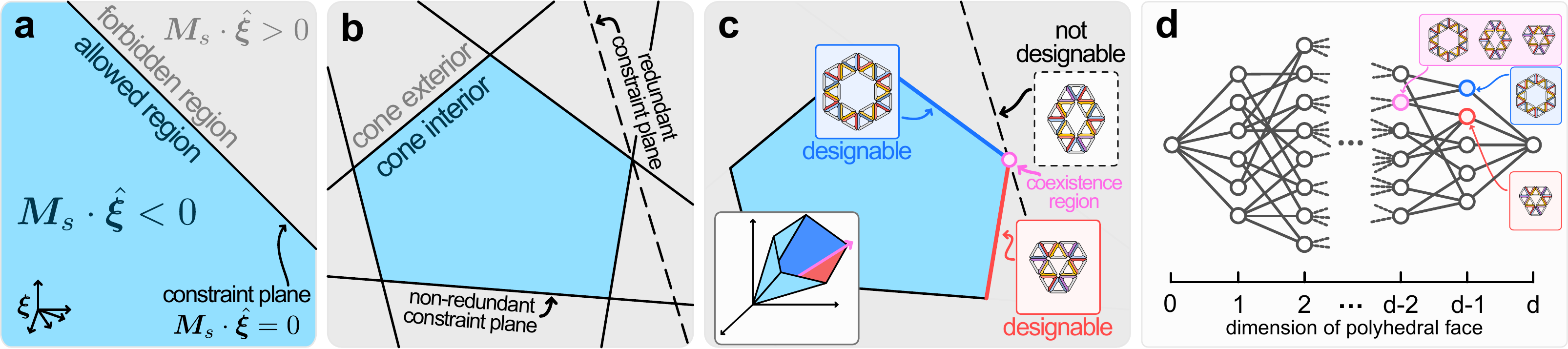}
    \caption{The polyhedral structure of the thermodynamic constraints. 
    (a) A constraint plane corresponding to a structure $s$, defined by $\M \cdot \hat\pars = 0$, separates the parameter space into two half-spaces. The gray half-space in which $\M \cdot \hat\pars > 0$ is physically forbidden in the limit of high $\lambda$, which restricts the allowed limiting directions to the space $\M \cdot \hat{\pars} \leq 0$.
    (b) The intersection of all allowed half-spaces forms a convex polyhedral cone. Redundant constraint planes may ``touch'' the cone, but the cone is unaffected by their presence. For clarity, this figure sketches a two-dimensional slice through a higher-dimensional space.
    (c) To every $d-1$-dimensional face of the constraint cone corresponds a designable structure.
    Lower dimensional faces correspond designable sets of structures. Structures corresponding to redundant constraints are not designable.
    A three-dimensional cartoon of the constraint cone is shown in the inset in the lower left.
    (d) The relationship between the faces of the constraint cone, and therefore the designable sets, can be visualized as a \emph{Hasse} diagram, as sketched in the cartoon example here. Nodes in the diagram correspond to faces/designable sets and edges indicate a containment relation: a $d_f$ dimensional set is contained in a $d_f - 1$ dimensional set if they are connected in the diagram.
    }\label{fig:theory}
\end{figure*}

We begin with a simple but far-reaching observation.
Achieving 100\% yield for a target structure $s_t$ requires a vanishing number density for all {\it other} structures, but since $\pars$ only appears in the exponent in Eq.~\eqref{eq:rho}, this can happen only when $\bm{M}_{s'} \cdot \pars \to -\infty$ for all structures $s' \neq s_t$. 
This means we must consider \emph{limits} in parameter space.
We thus rewrite $\pars$ as $\pars = \lambda \hat\pars$, where $\rVert \hat\pars \rVert = 1$, so we can systematically take the limit $\lambda\to\infty$. Note that this is a seemingly diabolical limit because as the binding energies diverge, so too do the equilibration times. Nevertheless, we will see that this limit implies thermodynamic constraints that have profound implications even as we pull back to experimentally relevant energy scales. 

In the limit of large $\lambda$, the density of any structure $s$ is given by
\begin{equation}\label{eq:limit}
    \lim_{\lambda \to \infty} \rho_s(\lambda \hat\pars) = \begin{cases}
        \infty  &\text{ if } \M \cdot \hat\pars > 0 \\
        \Omega_s  &\text{ if } \M \cdot \hat\pars = 0 \\
        0  &\text{ if } \M \cdot \hat\pars < 0 \,.
    \end{cases}
\end{equation}
Importantly, the first case, $\M \cdot \hat\pars > 0$, implies diverging particle concentrations, meaning that these limits cannot be physically realized and need to be excluded (see Methods).
Thus, for each structure $s$, there exists a thermodynamic constraint in the asymptotic limit given by 
\begin{equation}\label{eq:constraints}
    \M \cdot \hat\pars \leq 0 \,.
\end{equation}

These constraints have an interesting geometrical structure.
Each constraint slices the $\dim$-dimensional parameter space in half, as illustrated in Fig.~\ref{fig:theory}a. As $\lambda\to\infty$, the gray region is forbidden while $\rho_s$ vanishes in the blue region; only when $\hat\pars$ is placed on the $\dim-1$ dimensional constraint plane where $\M\cdot \hat\pars=0$ does the structure $s$ assemble at finite concentration in this limit. However, there are many such constraint planes -- one for every structure allowed by the binding rules. As illustrated in Fig.~\ref{fig:theory}b, these constraints work together to further restrict the region in parameter space that is allowed. Geometrically, this region forms a $\dim$-dimensional convex polyhedral cone, or constraint cone, whose boundary is composed of constraint planes.

This constraint cone allows us to understand exactly what can and cannot happen in the $\lambda\to \infty$ limit. Placing the parameters in the cone's interior (blue region) means that the number density of all structures goes to zero. But if we align the parameters with one of the $d-1$ dimensional faces of the cone, then the structure corresponding to that constraint plane -- and only this structure -- will assemble at finite number density and thus achieve 100\% yield. For example, placing $\hat\pars$ anywhere along the dark blue face in Fig.~\ref{fig:theory}c will assemble the hexagon structure at 100\% yield, while placing $\hat\pars$ on the red face will assemble the triangular structure at 100\% yield. We say that these two structures are \emph{designable}.

However, not every structure is designable. The constraint plane corresponding to the rhomboid structure is shown by the dashed line and only intersects the cone at the intersection of the dark blue and red faces (purple dot in Fig.~\ref{fig:theory}c).
If we place $\hat\pars$ at this intersection then all three structures will assemble with non-zero yield, meaning the rhomboid structure is not designable. 
Nevertheless, this observation allows us to expand the notion of designability to sets of structures that together can assemble at 100\% yield. The three structures shown in Fig.~\ref{fig:theory}c form such a \emph{designable set}, and others can be found by looking at similar intersections of the constraint planes. 

\begin{figure*}
\centering
\includegraphics[width=\textwidth]{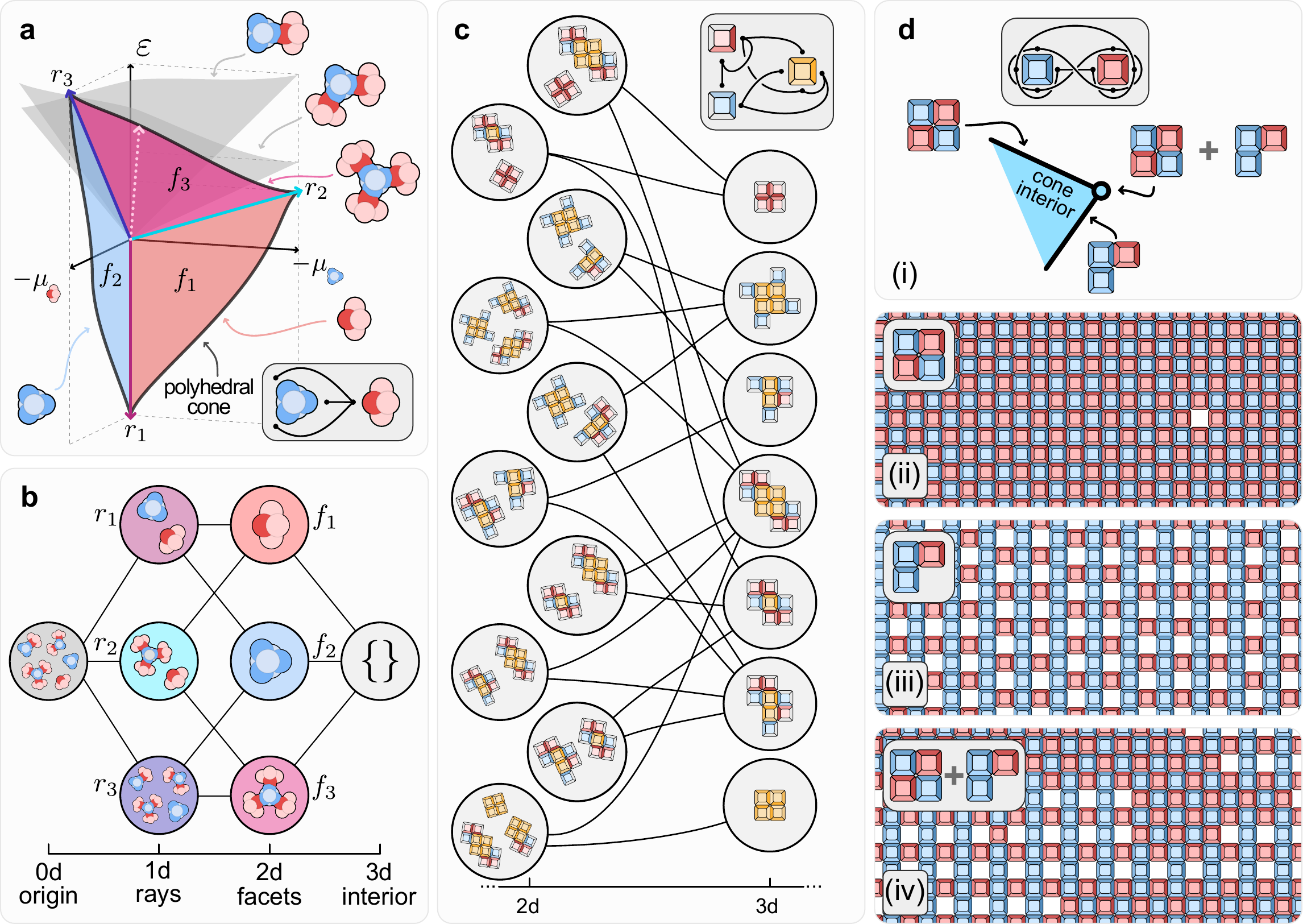}
    \caption{Physical implications of the polyhedral structure.
    (a) The constraint planes and polyhedral cone derived from the simple binding rules shown in the bottom right. The parameter space consists of the binding energy $\varepsilon$ (defined here to be positive for attractive interactions and the same for all bonds) and the chemical potentials $\bm\mu$ of the two particle species. Non-redundant constraints are shown in color and redundant constraints are shown in gray. The white dotted line shows a limit direction parallel to the face $f_3$.
    (b) The faces of the polyhedral cone in (a), visualized as a Hasse diagram. Every node corresponds to a polyhedral face, which in turn corresponds to a designable set of structures.  The edges indicate containment relations.
    (c) A section of the Hasse diagram corresponding to the complex, non-deterministic binding rules shown in the top right. Shown are only designable structures and designable sets that do not contain free monomers.
    (d) The binding rules shown on the top of (i) lead to a large number of possible crystal phases. (i) further shows a cartoon sketch of a high-dimensional intersection of faces in the approximate constraint cone, which predicts that the checkerboard tiling and the tiling with holes can coexist. (ii) Simulation snapshot of a checkerboard tiling. (iii) Simulation snapshot of a tiling with holes. (iv) Simulation snapshot of a coexistence of checkerboard and tiling with holes. See SI for details on the simulations.
    }
    \label{fig:examples}
\end{figure*}

The intersections of various high-dimensional constraint planes are more complicated than in the simple 2-dimensional illustration in Fig.~\ref{fig:theory}a-c. 
In a $\dim$ dimensional parameter space, the constraint planes define $\dim-1$ dimensional faces of the cone, while the intersection of two such faces forms a $\dim-2$ dimensional face. Furthermore, $\dim-2$ dimensional faces can intersect to form $\dim-3$ dimensional faces, and so on until you arrive at the $0$ dimensional ``face'' at the origin, $\pars=0$. 
Using tools from polyhedral computation (Methods), we can identify every face $f$ of the constraint cone, which we organize in a so-called {\it Hasse diagram} according to each face's dimensionality $\dim_f$ (Fig~\ref{fig:theory}d).
The faces of a convex polyhedron are always nested within each other, and the inclusion relations are visualized by the edges in the Hasse diagram.

The key insight is that each face, regardless of its dimensionality, corresponds to a set of structures that together are designable. These structures can be assembled at combined 100\% yield by aligning the parameters with the face and taking $\lambda \to\infty$.
Furthermore, this geometrical and combinatorial structure of the designable sets allows us to identify lower-dimensional design spaces -- unbreakable rules created by statistical mechanics that govern relative yields within these designable sets in thermal equilibrium.

To see this, choose an arbitrary face $f$. We can write any set of parameters $\pars$ as 
\begin{equation}
    \pars = \lambda \hat \pars_f + \pars_\perp,
\end{equation}
where $\lambda \hat\pars_f$ is the $\dim_f$ dimensional component of $\pars$ that is parallel to $f$ and $\pars_\perp$ is the $c_f\equiv\dim - \dim_f$ dimensional component that is perpendicular to $f$. 
For structures not in the corresponding designable set $\mathcal{S}_f$, the extra finite component $\pars_\perp$ does not affect the assembly as $\lambda \to\infty$ and the structures are still completely suppressed. For structures $s$ in $\mathcal{S}_f$, however, the number density becomes 
\begin{equation} \label{eq:rho_s_from_xi_perp}
    \rho_s = \Omega_s e^{\M\cdot \pars_\perp} \,,
\end{equation}
meaning that the relative yields within the set can be tuned by varying $\pars_\perp$.
In fact, Eq.~\eqref{eq:rho_s_from_xi_perp} is valid for any $\lambda$, even far away from the asymptotic limit. In other words, the number densities of structures within a designable set $\mathcal{S}_f$ are controlled by a $c_f$ dimensional space perpendicular to $f$. See SI for details. 

We will now explore these ideas through a series of examples that will highlight specific insights, consequences, and experimental implications. 

\subsection{Exploring the polyhedral structure through specific examples}
\paragraph{First, a minimal example}
We begin with a very simple example consisting of two particle species, each with their own chemical potential, that can bind as shown in Fig.~\ref{fig:examples}a.
Here, all possible bonds are of the same type, governed by a single binding energy $\varepsilon$, which, together with two chemical potentials, form a three-dimensional parameter space.
These binding rules allow five distinct structures to form: the blue and red monomers, a dimer, a trimer, and a tetramer.
Figure~\ref{fig:examples}a shows the five constraint planes, defined by $\M \cdot \hat\pars = 0$, from which considerable insight can be gained.

First, notice that the region that satisfies all constraints -- the constraint cone -- is only bounded by three of the five planes.
The corresponding three structures, both monomers and the tetramer, are therefore designable: each one can be assembled at high yield by aligning the parameters with the corresponding face ($f_1$, $f_2$, or $f_3$) and taking the asymptotic limit. 
In contrast, the dimer and trimer are not individually designable because their constraint planes (shown in gray) only touch the constraint cone at the lower-dimensional face $r_3$. Assembling either structure at high yield therefore requires aligning the parameters with $r_3$, but doing so makes it impossible to suppress the tetramer or blue monomer. No matter how energies or concentrations are chosen, the dimer and trimer can never assemble alone. Instead, these four structures together form a designable set, corresponding to the face $r_3$. 
The other two designable sets are the two monomers (corresponding to $r_1$) and the red monomer plus the tetramer (corresponding to $r_2$), see the Hasse diagram in Fig.~\ref{fig:examples}b. 

\paragraph{Reconfigurable assembly with complex binding rules}
We now apply our framework to the complex binding rules shown in Fig.~\ref{fig:examples}c, which were originally investigated in Ref.~\cite{Huebl.2024} as an economical design for a specific structure shape.
Here we show that these rules can do much more.
Assuming uniform binding energies (not required for our theory, but convenient in many experiments) leads to four degrees of freedom: three chemical potentials and one binding energy.
We enumerate all 677 possible structures using the tools of Ref.~\cite{Huebl.2024}, and construct the constraint cone. 

Investigating the constraint cone reveals that there are seven individually designable structures (excluding free monomers), which are shown in the partial Hasse diagram in Fig.~\ref{fig:examples}c.
In practice, this means that an experiment can control which one of these structures assembles simply by tuning particle concentrations, without having to redesign the interactions.
This is thus an example of reconfigurable assembly, where one set of binding energies can lead to different assembly outcomes, depending on external input. 
See SI for a detailed discussion.

Going further, the designable sets shown in the next (2d) level in the Hasse diagram tell us which of these seven structures can be assembled simultaneously.
Notice that some of these designable sets also include a third, non-designable structure: these are unavoidable chimeras that cannot be suppressed if the other two structures should assemble together.

\begin{figure*}
\centering
\includegraphics[width=\textwidth]{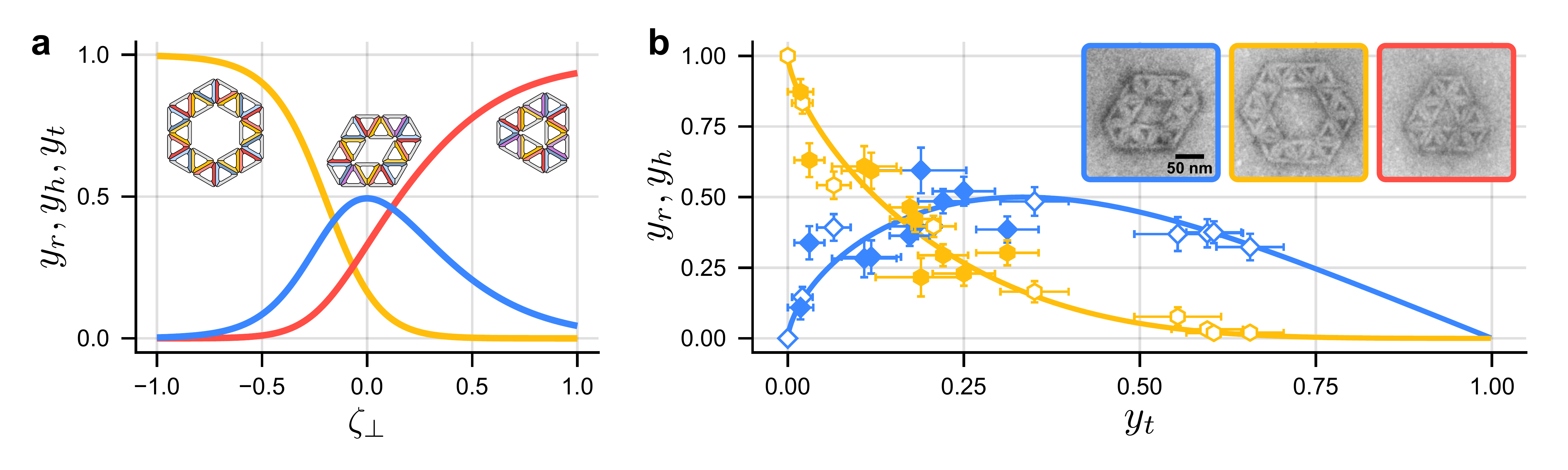}
    \caption{
    Quantitative experimental validation of the reduced design space. 
    (a) Relative yield space of rhomboid ($y_\mathrm{r}$, blue), hexagon ($y_\mathrm{h}$, yellow), and triangular ($y_\mathrm{t}$, orange) rings. Shown are the relative yields $y_s = Y_s / (Y_\mathrm{tri} + Y_\mathrm{rho} + Y_\mathrm{hex})$, as a function of the one degree of freedom $\zeta_\perp$ in parameter space that affects the relative yields.
    (b) Comparison between experimentally measured (points) and theoretically predicted (lines) relative yields of the three rings. 
    Since we do not know the value of $\zeta_\perp$ in experiments, $y_\mathrm{r}$ and $y_\mathrm{h}$ are shown as a function of $y_\mathrm{t}$. 
    Empty symbols correspond to equal interactions, full symbols correspond to enhanced binding between some bonds as defined in the text. Different data points are obtained at varying concentrations of the purple particle $c_\mathrm{p}$ (see Fig.~\ref{fig:intro}b) and MgCl$_2$ concentrations.
    Error bars show the standard error of the measured relative yields. See SI for details.}
    \label{fig:rings}
\end{figure*}

\paragraph{Coexisting crystals}
Constructing the polyhedral constraint cone requires a complete enumeration of all allowed structures. For binding rules that lead to crystallization or other bulk assemblies, such an enumeration is not possible. Nevertheless, we now show that nontrivial insight and predictions into such systems can be derived through an \textit{incomplete} enumeration. 
To this end, we consider the binding rules shown in Fig.~\ref{fig:examples}d(i), consisting of two particle species and 8 independently tunable binding energies.

As described in the SI, we construct an approximate constraint cone by sampling small clusters and periodic unit cells, and identify 77 that are designable.
Two of these designable unit cells are shown in Fig.~\ref{fig:examples}d(i). Our theory thus predicts that high-quality crystals, with these unit cells, can be achieved by placing the parameters along the corresponding faces and taking the $\lambda \to \infty$ limit. To verify this, we simulate crystal formation using a simple grand-canonical Monte Carlo scheme similar to the one used in Ref.~\cite{Murugan.2015kjd} (see SI).
Figures~\ref{fig:examples}d(ii-iii) show sections of the resulting crystal phases, confirming high-quality crystallization with only minimal defects. 

Furthermore, the constraint planes of the two tilings intersect in a lower-dimensional face without any unavoidable chimeras. Our theory thus predicts parameter values at which coexistence between these two bulk phases is possible, which we confirm through simulation, as shown in Fig.~\ref{fig:examples}d(iv), and with all details in the SI.
While this is only a first step towards a full treatment of phase coexistence through the present framework, this example highlights the applicability of our theory to bulk condensation, and shows that a complete enumeration of all structures is not always required for the constraint cone to make accurate predictions.

\paragraph{Coexistence of ring-shaped structures}
We now return to our introductory example shown in Fig.~\ref{fig:intro}b, and analyze it using our developed framework.
The four particle species interact with five bond types, which, assuming rigid and in-plane assembly, lead to 283 different possible structures.
Out of these, we focus specifically on the three closed rings (Fig.~\ref{fig:intro}c(i)).
We view all five binding energies as free parameters, leading to a $d=9$ dimensional parameter space.

Constructing the constraint cone reveals that there are 21 individually designable structures, which include the hexagonal ring and the triangular ring.
However, the rhombus is \emph{not} individually designable, but can only be assembled as part of a larger set that contains all three rings together.
This situation was already sketched qualitatively in Fig.~\ref{fig:theory}c: the constraint corresponding to the rhomboid is redundant and only touches the constraint cone at a $d_f=7$ dimensional face.

To understand the possible relative yields between the three rings, we follow our earlier discussion and decompose parameter space into a subspace parallel to the face associated with the designable set, and a subspace orthogonal to it.
The orthogonal subspace governs relative yields, and is $c_f = 2$ dimensional in this case.
However, in this specific example, one of these two degrees of freedom only leads to a uniform rescaling of the overall number density (see Methods), which leaves only a single degree of freedom to tune the relative yields between the three rings.

Figure~\ref{fig:rings}a shows the relative yields of the three rings as a function of the relevant $1$-dimensional component of $\pars$, which we label $\zeta_\perp$. No matter how the energies and particle concentrations are chosen, thermodynamically allowed relative yields must follow these curves, whose shapes are determined by structure compositions and entropy (Methods). 

\subsection*{Experiments corroborate the polyhedral structure}
We now introduce an experimental system to explore the real-world implications of our theory.
We synthesize triangular particles using DNA origami, and implement the specific binding rules shown in Fig.~\ref{fig:intro}a by extending single-stranded DNA from the triangles' sides.
Adjusting the sequences of these strands then allows us to program specific interactions by exploiting Watson-Crick base pairing, and folding monomers with various combinations of edge strands allows us to create different ``colors'' of particles, or species (see SI for sequence information). We then perform assembly experiments by mixing the various particle types with concentrations between 10~nM and 50~nM for an individual particle type in a buffer that has 20--30~mM of MgCl$_2$. By annealing these assembly mixtures from 40--25~$^\circ$C with a ramp rate of 0.1~$^\circ$C$/1.5$~h, we ensure that we pass through the temperature at which the interactions are weak and reversible, and the structures can anneal. After annealing is complete, we image the assembly result using transmission electron microscopy (TEM), allowing us to count different structure shapes to measure their relative yields.

Since the particles assemble in three dimensions with finite out-of-plane flexibility, there are many other possible off-target structures not taken into account in our enumeration (Fig.~\ref{fig:intro}c).
We thus expect the absolute yield of the target structures to remain low, especially since the experimentally achievable energy scales are not high enough to reach the asymptotic limit, where complete suppression of all off-target structures becomes possible.
On the other hand, our prediction that there is only a single degree controlling the relative yields of the three target structures is independent of the number of off-target structures or the asymptotic limit, and thus constitutes a robust prediction that we now test.

To this end, we perform multiple experiments at different parameter values, changing the concentration of purple particles $c_\mathrm{p}$ (see Fig.~\ref{fig:intro}b(i)) and the relative binding energies between particles. 
Two sets of binding energies were used, one where all interactions are equally strong, and another where the purple--blue and purple--yellow interactions are weaker than the other interactions (see SI for details).
Since the value of $\zeta_\perp$ (see Fig.~\ref{fig:rings}a) is unknown in experiments, we re-plot the different yield curves of Fig.~\ref{fig:rings}a against each other, showing the normalized yields of the hexagonal and rhomboidal rings as a function of the triangular ring. 
In this way, we can directly compare our theoretical results to experimental data.

Figure~\ref{fig:rings}b shows excellent agreement between our theoretical expectations and experimental results. 
The lines show the theoretically predicted relative yields, whereas the experimentally measured relative yields are shown with different markers, according to the experimental parameter values.
Every data point corresponds to between 50--150 countable structures.
This striking quantitative agreement between theory and experiment, which is achieved without any fit parameters and with minimal assumptions on the particle interactions (see Methods), confirms that the polyhedral structure of the thermodynamic constraints leads to robust experimental predictions.

Moreover, this agreement is obtained very far from the asymptotic limit, with absolute yields of the rings ranging from about 0.1--1.2\% (corresponding to 1.3--18\% mass-weighed yield, see SI). This confirms the key finding of the paper: equilibrium self-assembly is governed by a polyhedral structure that restricts which yield combinations are allowed. For example, even with 9 parameters, the three yields in Fig.~\ref{fig:rings} cannot be tuned independently. Our ability to not only understand but also precisely calculate this underlying polyhedral structure makes it a powerful tool for experimentally accurate inverse design.

\subsection*{Discussion}
While the majority of work in programmable self-assembly focuses on tuning the binding rules, we have shown how to comprehensively understand the secondary design space consisting of binding energies and chemical potentials. For a given choice of binding rules, thermodynamic constraints form a high-dimensional polyhedral cone, and the faces of this cone determine which sets of structures are designable, {\it i.e.} can theoretically assemble at high yield.
Moreover, this framework identifies best-case scenarios for non-designable structures, necessary chimeras, low-dimensional design spaces, and reconfigurable assemblies, thus providing a robust and powerful tool for economical inverse design. 

The excellent experimental agreement that we find in Fig.~\ref{fig:rings} both validates a key pillar of our theory and demonstrates its utility in practice. However, assembly outcomes that are theoretically possible may still be challenging to realize experimentally, for example due to long equilibration times, or experimental bounds on achievable parameter values.
The continued development of experimental capabilities
will help to achieve these challenging high-yield assemblies.
In addition, despite the success shown in Fig.~\ref{fig:examples}d,
our framework rests on the ability to identify important competing structures, and while the tools of Ref.~\cite{Huebl.2024} are sufficient for our purposes, further work is needed to identify structures with internal stress or more robustly deal with bulk assemblies.

Our theory is derived from a single assumption, that the free energy of a self-assembled structure is linear in the relevant design variables. In practice, this holds for binding energies and chemical potentials in systems assembling with short-ranged interactions.
This form of the free energy is routinely encountered in self-assembly~\cite{Israelachvili2011}, and, besides DNA origami, has been used to describe systems as diverse as DNA-coated colloids~\cite{Valignat.2005, Gruenwald.2014, McMullen.2022, Holmes-Cerfon.2016}, colloids with depletion interactions~\cite{Meng.2010, Sacanna.2010, Perry.2016}, virus capsids and other protein complexes~\cite{Zandi.2006, Perlmutter.2015, Zlotnick.2005, Sartori.2020, Curatolo.2023}, and magnetic-handshake particles~\cite{Du.2022}, among others.
This broad generality comes from the fact that the system-specific details of the binding interactions affect the structure entropy $\Omega_s$, but leave the essential scaling of the Boltzmann weight, $e^{\M \cdot \pars}$, unaffected.
This means that the designability of (sets of) structures is independent of system details, while the precise shape of relative yield curves, such as those in Fig.~\ref{fig:rings}, may change depending on the details of the binding interactions.
In our specific example in Fig.~\ref{fig:rings}, we were able to predict the shape of the relative yield curves without modeling the microscopic interactions, because the shared ring-like topology of the structures causes the structure entropies to drop out of the relative yield calculations (see Methods).

To see how our framework might be used in practice, start with one or more target structures with the goal of assembling them at high yield. 
For a proposed set of binding rules (which could be generated with the methods of Refs.~\cite{Bohlin.2023, Huebl.2024}, for example), our calculations enable a comprehensive and near-instantaneous view of the secondary design space. 
We simply read out whether or not the target structures are designable, allowing us to quickly reject binding rules for which the best-case yield is insufficiently small. For example, we can immediately reject the binding rules in Fig.~\ref{fig:intro}b if we want to assemble the rhomboid on its own. However, these binding rules are excellent if our goal is to dynamically switch between the hexagon and triangle, as this can be achieved by moving along the low-dimensional design space $\zeta_\perp$. 

Our results raise a number of important and interesting questions. First and foremost, does self-assembly far from equilibrium conform to similar constraints? While counterexamples surely exist, it will be interesting to learn whether the polyhedral structure provides a useful starting point for understanding and designing non-equilibrium systems. Furthermore, has evolution exploited the polyhedral structure to better achieve robust and economical assembly?
A better understanding of the dynamical processes that govern the approach to equilibrium or the drive away from equilibrium will help to address these questions more thoroughly, opening new doors to the design and control of dissipative pathways to synthetic assembly, and a better understanding of self-organization in living systems. 

\section*{Methods}
\subsection*{Yield calculations and their experimental validation}
As discussed in the main text, we consider design parameters consisting of the chemical potentials $\mu_\alpha$ of each particle species $\alpha$ and the binding energies $\varepsilon_{\alpha\beta, ij}$ between sites $i$ and $j$ of particle species $\alpha$ and $\beta$, where a positive $\varepsilon_{\alpha\beta, ij}$ indicates an attractive interaction.
To simplify the notation, we collect the chemical potentials and nonzero binding energies in a parameter vector
\begin{equation}
    \pars = \frac{1}{k_\mathrm{B}T}\left[\begin{array}{c}
        \mu_1  \\
        \vdots \\
        \mu_{n_\mathrm{spc}} \\
        \varepsilon_{\alpha_1 \beta_1, i_1 j_1} \\
        \vdots \\
        \varepsilon_{\alpha_{n_{\mathrm bnd}} \beta_{n_{\mathrm bnd}}, i_{n_{\mathrm bnd}} j_{n_{\mathrm bnd}}}
    \end{array}
    \right]
\end{equation}
where $n_\mathrm{spc}$ is the number of particle species and where the $n_\mathrm{bnd}$ bonds with nonzero binding energy are defined by the binding rules (e.g., Fig.~\ref{fig:intro}b).
$\pars$ has dimension $d =n_\mathrm{spc} + n_\mathrm{bnd}$.
The vector $\M$ that appears in Eq.~\eqref{eq:rho} counts the number of species and bond types present in structure $s$.
If the structure contains $n^\alpha_s$ particles of species $\alpha$, and $b^{\alpha\beta, ij}_s$ bonds connecting binding sites $i$ and $j$ of species $\alpha$ and $\beta$, then 
\begin{equation}
    \M = \left[\begin{array}{c}
        n^1_s  \\
        \vdots \\
        n^{n_\mathrm{spc}}_s \\
        b^{\alpha_1 \beta_1, i_1 j_1}_s \\
        \vdots \\
        b^{\alpha_{n_{\mathrm bnd}} \beta_{n_{\mathrm bnd}}, i_{n_{\mathrm bnd}} j_{n_{\mathrm bnd}}}_s
    \end{array}
    \right]
\end{equation}

The pre-factor $\Omega_s$ is the entropic partition function of a structure $s$, and is given by
\begin{equation}
    \Omega_s = \frac{Z^\mathrm{vib}_s \, Z^\mathrm{rot}_s}{\lambda^{D n_s} \sigma_s } \,,
\end{equation}
where $Z^\mathrm{vib}_s$ and $Z^\mathrm{rot}_s$ are the vibrational and rotational partition functions~\cite{Klein.2018, Curatolo.2023} of $s$ respectively, $\lambda^D$ is the volume of a phase space cell, and $n_s = \sum_\alpha n^\alpha_s$ is the number of particles in $s$.
The symmetry number $\sigma_s$ counts the number of permutations of particles in $s$ that are equivalent to a rotation of the structure~\cite{Klein.2018}.
The multiplicity coming from permutations of identical particle species is taken into account implicitly through the chemical potentials~\cite{Huebl.2024, Curatolo.2023, Kardar.2007}.
If the microscopic interactions between particles are known, $\Omega_s$ can be computed using the methods in Refs.~\cite{Klein.2018, Curatolo.2023, Holmes-Cerfon.2013, Holmes-Cerfon.2016}.

We assume $\Omega_s$ to be independent of $\pars$, which is exactly true for simple interaction models, such as those in Ref.~\cite{Huebl.2024}, and a good assumption in general, since the effect of the parameters on structure entropy is much weaker than their contribution in the Boltzmann weight, Eq.~\eqref{eq:rho}.
For the DNA-origami particles that we study here, models of the DNA-mediated binding interactions suggest that the binding entropy and energy can, to some extent, be tuned independently~\cite{Videbaek.2025}.

When enumerating structures in this paper, we assume that bond stiffness is very high, such that building blocks can only fit together side-to-side and strained structures cannot form.
This assumption could be relaxed~\cite{Huebl.2024}, in which case the partition function for a strained structure would carry an additional factor $e^{-\varepsilon_\mathrm{strain}/kT}$, set by the strain energy $\varepsilon_\mathrm{strain}$.
As long as this strain energy does not strongly depend on, or is linear in the design parameters, our framework applies to strained structures as well.

Figure~\ref{fig:intro}d shows that equilibrium yields, as predicted by Eq.~\eqref{eq:yields}, agree excellently with the experimental yields.
We have realized the particles and binding rules shown on the left-hand side of Fig.~\ref{fig:intro}d with DNA origami (the details of which are discussed below and in the SI), and we measured the yields of the seven possible structure shapes as a function of the concentration $c_\mathrm{red}$ of the particle species shown in red.
All other particle species were kept at $c_\mathrm{blue} = c_\mathrm{yellow} = c_\mathrm{purple} = c_0$, with $c_0 = 2~\mathrm{nM}$.
The binding energies of all four bond types were designed to be the same in the experiments.
Because the precise value of the binding energy is unknown, we use it as a fit parameter in our theoretical calculations.
Minimizing the least-squares error between the yields predicted by Eq.~\eqref{eq:yields} and the experimental yields, we obtain optimal agreement at a fit binding energy of approximately $12.1 \, k_{\rm B}T$. This value is in line with previous experimental estimates of the binding energy for DNA-origami triangles~\cite{Hayakawa.2022pqs, Wei.2024}.

\subsection*{Structure enumeration}
Most of the results in this paper were obtained with help from the structure enumeration algorithm introduced in Ref.~\cite{Huebl.2024}, which is capable of efficiently generating structures in 2d or 3d that satisfy a given set of binding rules, assuming rigidly-locking binding interactions.
The algorithm enumerates all physically realizable structures that can be formed from given binding rules, meaning that steric overlaps and bonds between incompatible binding sites are forbidden.
Particle overlaps are easily detected for particles assembling on a lattice (as is the case in all systems shown here); more complex building blocks can, in principle, be represented as rigid clusters of spheres or triangulated meshes to find overlaps and contacts~\cite{Huebl.2024}.

The algorithm can enumerate roughly 10,000 structures per second, and all enumerations in this paper were performed in less than one second on a 2024 MacBook Pro.
The algorithm makes it possible to quickly detect whether a given set of binding rules leads to infinitely many structures or not, and can optionally generate structures only up to a maximal size.
It is, in principle, possible to extend the enumeration algorithm to enumerate structures with flexible bonds.
This would require a model of the microscopic binding interactions between particles, and would lead to additional computational costs~\cite{Huebl.2024}.

It is important to note that our theory is completely agnostic towards what enumeration method is used, and in general, the best (\emph{i.e.}, most efficient or convenient) method for generating structures depends on the system at hand.
For example, very different algorithms have been used to enumerate rigid sphere clusters Refs~\cite{arkus.2009, Arkus.2011, Holmes-Cerfon.2016}, or conformations of colloidal polymers~\cite{McMullen.2022}.

\subsection*{Thermodynamic constraints}
To see why diverging structure densities in the asymptotic limit are unphysical, note that for parameters violating $\M \cdot \hat\pars \leq 0$, particle concentrations \emph{rise} as $\lambda$ is increased, meaning that to realize the asymptotic limit, more and more particles need to be added to the system.
At some point, steric effects, which are not explicitly modeled here, make it impossible to add more particles, which means that chemical potentials cannot be raised further, making it impossible to reach the asymptotic limit.

This situation is similar to, but distinct from, other cases where unphysical chemical potentials emerge, such as standard aggregation theory~\cite{Hagan.2021}, or \textit{e.g.} degenerate Bose gases~\cite{Kardar.2007}.
In these cases, unphysical chemical potentials arise due to singularities in the partition function, while in our case (assuming a finite number of possible structures) the forbidden chemical potentials come from the imposition of the asymptotic limit in parameter space.

\subsection{Designable sets and polyhedral faces}
We now give more precise definitions to the concepts discussed in the main text.
All of the following definitions are discussed at length in any reference on convex polyhedra, such as Refs.~\cite{Ziegler.1995, Gruenbaum.2003, Barvinok.2002, fukuda2020}.

Consider a polyhedron defined by a series of linear inequality constraints, Eq.~\eqref{eq:constraints}, with $s \in \mathcal{S}$.
A polyhedral face $f$ is a subset of the polyhedron where certain constraints are active ($\M\cdot\hat\pars = 0$ for $s \in \mathcal{S}_f \subset \mathcal{S}$), and all other constraints are inactive ($\M\cdot\hat\pars < 0$ for $s \notin \mathcal{S}_f$).
The faces of a polyhedral cone can have any dimension from $d_f = 0$ to $d_f = d$.
Comparing this definition with the definition of designable sets, Eq.~\eqref{eq:limit}, shows that the directions in parameter spaces that lead to high-yield assembly for a designable set $\mathcal{S}_f$ correspond one-to-one with the polyhedral face $f$ whose active equalities correspond to the structures in $\mathcal{S}_f$.

Faces of a polyhedron can be partially ordered by inclusion, and the resulting partially ordered set is called the polyhedron's \emph{face lattice}.
The combinatorial properties of the face lattice give rise to the combinatorial properties of designable sets, which are visualized with the Hasse diagrams in the main text.

Perhaps the most important combinatorial property of designable sets is that the intersection of two designable sets is again designable.
This is proved in SI, and has an important consequence: for any group of structures, there exists a unique smallest designable set that contains all of them.
Optimizing the yield of a structure that is not designable by itself therefore consists of finding this minimal compassing designable set, and then tuning relative yields to maximize the yield of the structure as much as possible, even if 100\% yield is impossible.

Furthermore, the polyhedral structure also has important computational implications.
Since all constraints are linear, checking whether a (set of) structure(s) is asymptotically designable can be done exactly and efficiently (in polynomial time in the number of constraints) with linear programming~\cite{boyd2004convex}, while the enumeration of designable sets can be achieved with algorithms from polyhedral computation, as shown below.

\subsection*{Polyhedral computation}
We compute the Hasse diagram of the constraint cones with a simple version of the algorithm described in~\cite{Fukuda.1994}. 
Briefly, we start by filtering the constraints for redundancies using \emph{cddlib}~\cite{CDDLib}, which leaves us with all non-redundant constraints.
We then use the double description algorithm~\cite{Motzkin1953, fukuda1996dd} to transform the cone from its inequality representation to its ray (vertex) representation.
From this, we can construct an incidence matrix that indicates which rays are contained in which facets of the polyhedron.
By generating the unique Boolean products of the columns of this matrix, we can iteratively construct all faces of the polyhedron.
This process takes less than a second for all systems considered here.

While the computational cost of computing the entire Hasse diagram scales exponentially with the number of non-redundant constraints and therefore becomes infeasible for large systems, it is important to note that the diagram can be generated layer by layer. This means that the ``right-most'' layers of the Hasse diagram, corresponding to the individually designable structures and small designable sets, can always be done rather quickly (more precisely, in polynomial time ) simply through \emph{redundancy removal} of the constraint~\cite{fukuda2020, clarkson1994, Avis.2024}.

\subsection*{Predicting allowed relative yields}
To find the direction in parameter space that allows tuning of the relative yields in Fig.~\ref{fig:rings}, we construct the matrix $M_\mathrm{rings} = [\bm{M}_\mathrm{tri},\bm{M}_\mathrm{hex}, \bm{M}_\mathrm{rho}]^T$.
Computing the singular value decomposition of this matrix shows that it has rank two, equal to the codimension $c_f$ of the corresponding face of the polyhedral constraint cone, as expected.
Importantly, however, the vector $\mathbbm{1} = [1, 1, 1]^T$ lies in the image of $M_\mathrm{rings}$, which means that there exists a direction in parameter space that raises the densities of the three rings uniformly and therefore does not affect relative yields.
Projecting this direction out leaves us with a rank-one matrix whose right singular vector corresponding to the non-zero singular value is the direction in parameter space that controls the relative yields between the three rings.

We do not model the microscopic interaction between particles, and thus cannot exactly calculate the value of $\Omega_s$ for the three rings.
We therefore estimate the rings' entropic partition functions as
\begin{equation}
    \Omega_s^\mathrm{rings} \approx \frac{4\pi v_\mathrm{eff}^{n_s-1}}{\lambda^{Dn_s}\sigma_s} \,,
\end{equation}
where $v_\mathrm{eff}$ is the effective volume a bound particle can explore if all other particles are held fixed and $n_s$ is the number of particles in structure $s$.
Importantly, this effective volume can be absorbed into the binding energies $\bm\varepsilon$, making them effective binding free energies $\tilde{\bm\varepsilon} = \bm\varepsilon + kT \log (v_\mathrm{eff} / \lambda^D)$.
This shows that $v_\mathrm{eff}$ can be compensated by changing $\bm\varepsilon$, which means that it cannot affect the range of possible relative yields and that we do not need to estimate $v_\mathrm{eff}$ for the predictions we make in the main text.
This is not true in general, but works in our case because the number of particles equals the number of bonds for all rings, causing $v_\mathrm{eff}$ to factor out.
Therefore, in this approximation, the shapes of the relative yield curves are determined by the structure compositions and the symmetry numbers $\sigma_\mathrm{hex} = 6$, $\sigma_\mathrm{tri} = 3$, $\sigma_\mathrm{rho} = 2$.

\subsection*{Folding DNA origami} To assemble our DNA origami monomers, we make a solution with 50~nM of p8064 scaffold (Tilibit), 200~nM of each staple strand (Integrated DNA Technologies [IDT]; Nanobase structure 247~\cite{Nanobase} for sequences), and 1x folding buffer. We then anneal this solution using a temperature protocol described below. Our folding buffer, from here on referred to as FoB\textit{X}, contains 5~mM  Tris Base, 1~mM EDTA, 5~mM NaCl, and \textit{X}~mM MgCl$_2$. We use a Tetrad (Bio-Rad) thermocycler to anneal our samples.

To find the best folding conditions for each sample, we follow a standard screening procedure to search multiple MgCl$_2$ concentrations and temperature ranges~\cite{Hayakawa.2022pqs,Wagenbauer2017}, and select the protocol that optimizes the yield of monomers while limiting the number of aggregates that form. All particles used in this study were folded at 17.5~mM MgCl$_2$ with the following annealing protocol: (i) hold the sample at 65~$^\circ$C for 15 minutes,  (ii) ramp the temperature from 58~$^\circ$C to 50~$^\circ$C with steps of 1~$^\circ$C per hour, (iii) hold at 50~$^\circ$C until the sample can be removed for further processing. \\

\subsection*{Agarose gel electrophoresis} We use agarose gel electrophoresis to assess the folding protocols and purify our samples with gel extraction. We prepare all gels by bringing a solution of 1.5\%~(w/w) agarose in 0.5X TBE to a boil in a microwave. Once the solution is homogenous, we cool it to 60~$^\circ$C using a water bath. We then add MgCl$_2$ and SYBR-safe (Invitrogen) to have concentrations of 5.5~mM MgCl$_2$ and 0.5x SYBR-safe. We pour the solution into an Owl B2 gel cast and add gel combs (20 $\upmu$L wells for screening folding conditions or 200~$\upmu$L wells for gel extraction), which cools to room temperature. A buffer solution of 0.5x TBE and 5.5~mM MgCl$_2$, chilled at 4~$^\circ$C for an hour, is poured into the gel box. Agarose gel electrophoresis is run at 110 V for 1.5--2 hours in a 4~$^\circ$C cold room. We scan the gel with a Typhoon FLA 9500 laser scanner (GE Healthcare) at 100~$\upmu$m resolution. \\

\subsection*{Sample purification} After folding, we purify our DNA origami particles to remove all excess staples and misfolded aggregates using gel purification. If the particles have self-complementary interactions, they are diluted 2:1 with 1xFoB2 and held at 47~$^\circ$C for 30 minutes to unbind higher-order assemblies. The folded particles are run through an agarose gel (now at a 1xSYBR-safe concentration for visualization) using a custom gel comb, which can hold around 2~mL of solution per gel. We use a blue fluorescent light table to identify the gel band containing the monomers. The monomer band is then extracted using a razor blade. We place the gel slices into a Freeze ’N Squeeze spin column (Bio-Rad), freeze it in a -20~$^\circ$C freezer for 5 minutes, and then spin the solution down for 5~minutes at 12~krcf. The concentration of the DNA origami particles in the subnatant is measured using a Nanodrop (Thermo Scientific). We assume that the solution consists only of monomers, where each monomer has 8064 base pairs.

Since the concentration of particles obtained after gel purification is typically not high enough for assembly, we concentrate the solution using ultrafiltration~\cite{Wagenbauer2017}. First, a 0.5-mL Amicon 100-kDa ultrafiltration spin column (Millipore) is equilibrated by centrifuging down 0.5~mL of 1xFoB5 buffer at 5~krcf for 7~minutes. Then, the DNA-origami solution is added and centrifuged at 14 krcf for 15 minutes. We remove the flow-through and repeat the process until all of the DNA origami solution is filtered. Finally, we flip the filter upside down into a new Amicon tube and spin down the solution at 1~krcf for 2~minutes. The concentration of the final DNA-origami solution is then measured using a Nanodrop. \\

\subsection*{Assembly experiments} Assembly experiments are conducted with DNA-origami particle concentrations ranging from 10~nM to 50~nM for the ring experiments (Fig.~\ref{fig:rings}b), and 6~nM to 10.5~nM for the small cluster experiments (Fig.~\ref{fig:intro}c). Assembly solutions have volumes up to 30~$\upmu$L with the desired DNA origami concentration in a 1xFoB buffer with MgCl$_2$ concentrations of 20~mM to 30~mM for the ring experiments and 20~mM for the small cluster experiments. During small cluster experiments, the solution is kept at room temperature. For ring experiments, the solution is placed in a 200~$\upmu$L PCR tube and loaded into a thermocycler (Bio-Rad), which is placed through a temperature ramp between 40~$^\circ$C and 25~$^\circ$C. The thermocycler lid is held at 100~$^\circ$C to prevent condensation of water on the cap of the PCR tube.

\subsection*{Negative-stain TEM} We first prepare a solution of uranyl formate (UFo). We boil doubly distilled water to deoxygenate it and then mix in UFo powder to create a 2\%~(w/w) UFo solution. We cover the solution with aluminum foil to avoid light exposure and vortex it vigorously for 20~minutes, after which we filter the solution with a 0.2~$\upmu$m filter. Lastly, we divide the solution into 0.2~mL aliquots, which are stored in a -80 $^\circ$C freezer until further use.

Before each negative-stain TEM experiment, we take a 0.2~mL UFo aliquot out from the freezer to thaw at room temperature. We add 4~$\upmu$L of 1~M NaOH and vortex the solution vigorously for 15~seconds. The solution is centrifuged at 4~$^\circ$C and 16~krcf for 8~minutes. We extract 170~$\upmu$L of the supernatant for staining and discard the rest.

The EM samples are prepared using FCF400-Cu grids (Electron Microscopy Sciences). We glow discharge the grid prior to use at -20~mA for 30 seconds at 0.1 mbar, using a Quorum Emitech K100X glow discharger. We place 4 $\upmu$L of the sample on the carbon side of the grid for 1 minute to allow adsorption of the sample to the grid. During this time, 5 $\upmu$L and 18 $\upmu$L droplets of UFo solution are placed on a piece of parafilm. After the adsorption period, the remaining sample solution is blotted on 11 $\upmu$m Whatman filter paper. We then touch the carbon side of the grid to the 5 $\upmu$L drop and blot it away immediately to wash away any buffer solution from the grid. This step is followed by picking up the 18 $\upmu$L UFo drop onto the carbon side of the grid and letting it rest for 30 seconds to deposit the stain. The UFo solution is then blotted and any excess fluid is vacuumed away. Grids are allowed to dry for a minimum of 15 minutes before insertion into the TEM. 

We image the grids using an FEI Morgagni TEM operated at 80 kV with a Nanosprint5 CMOS camera (AMT). The microscope is operated at 80 kV and images are acquired between x3,500 to x5,600 magnification. \\

\backmatter
\bmhead{Supplementary information}
Supplementary information, containing supplementary notes and examples, together with the description of numerical and experimental details, is available at [URL].

\bmhead{Acknowledgements}
We thank Berith Isaac and Amanda Tiano for their technical support with electron microscopy and Scott Waitukaitis for helpful comments on the manuscript. TEM images were prepared and imaged at the Brandeis Electron Microscopy facility. M.C.H and C.P.G acknowledge funding by the Gesellschaft f\"ur Forschungsf\"orderung Nieder\"osterreich under project FTI23-G-011. T.E.V, D.H., and W.B.R. acknowledge support from the Brandeis University Materials Research Science and Engineering Center (MRSEC) under grant number NSF DMR-2011846. W.B.R. acknowledges support from the Smith Family Foundation.

\section*{Declarations}
\subsection*{Conflict of interest}
The authors do not declare any conflict of interest.
\subsection*{Code availability}
Structure enumeration was performed with the \emph{Roly.jl}~\cite{Huebl.2024, Roly.jl} package developed by M.C.H and C.P.G, which is available at \url{https://github.com/mxhbl/Roly.jl}.
Polyhedral computation and linear programming were performed using the freely available \emph{Convex.jl}~\cite{convexjl}, \emph{Polyhedra.jl}~\cite{legat2023polyhedral} packages and \emph{cddlib}~\cite{CDDLib} library.
Example code reproducing the calculations done on the three rings and reconfigurable squares is available on GitHub at \url{https://github.com/mxhbl/PolyhedralStructureOfSelfAssembly}.
\subsection*{Data availability}
Design files and folding conditions of DNA origami used in this work can be found in the repository Nanobase~\cite{Nanobase} and are accessible at \url{https://nanobase.org/structure/247}.
\subsection*{Author contribution}
M.C.H. and C.P.G. conceived of and developed the theory. M.C.H. performed the numerical calculations. T.E.V., D.H., and W.B.R. designed the experiments. T.E.V. and D.H. conducted the experiments. All authors analyzed the data and wrote the paper.

\subsection*{Materials availability}
Not applicable

\end{document}


\title{Supplementary Information: The polyhedral structure underlying programmable self-assembly}

\author[1]{\fnm{Maximilian C.} \sur{H\"ubl}}\email{maximilian.huebl@ist.ac.at}
\author[2]{\fnm{Thomas E.} \sur{Videb{\ae}k}}\email{videbaek@brandeis.edu}
\author[2]{\fnm{Daichi} \sur{Hayakawa}}\email{dhayakawa@brandeis.edu}
\author*[2]{\fnm{W. Benjamin} \sur{Rogers}}\email{wrogers@brandeis.edu}
\author*[1]{\fnm{Carl P.} \sur{Goodrich}}\email{carl.goodrich@ist.ac.at}

\affil[1]{\orgname{Institute of Science and Technology Austria (ISTA)}, \orgaddress{\street{Am Campus 1}, \city{Klosterneuburg}, \postcode{3400}, \country{Austria}}}
\affil[2]{\orgdiv{Martin A. Fisher School of Physics}, \orgname{Brandeis University}, \orgaddress{\city{Waltham}, \postcode{02453}, \state{Massachusetts}, \country{USA}}}

\maketitle
\section*{Supplementary Methods}
\subsection{How to optimize the design parameters in practice}
In the main text, we mostly focus on polyhedral structure of the thermodynamic constraints, which quickly tell us what assembly outcomes are possible in principle, and which are not.
Here we go into more detail about \emph{how} these outcomes can be achieved, i.e. how to optimize the design parameters to achieve the desired assembly outcomes.

The polyhedral theory discussed in the main text allows us to obtain desired assembly outcomes simply by choosing the parameters $\pars$ to lie within a given face of the constraint cone.
This amounts to solving a system of linear inequalities (one for each non-redundant structure), which can be done efficiently with linear programming.
Moreover, since the solution is generally degenerate, we can use linear programming to optimize the rate of convergence, i.e. find the direction in parameter space that optimizes the speed at which undesired yield is suppressed as $\lambda$ increases. 

To find this optimal limiting direction $\hat \pars$, we minimize the largest element of $\bm{M}_s \cdot \hat{\pars}$ that does not correspond to a structure in the set whose yield we want to maximize:
\begin{align}\label{eq:minimize}
    \min_{\hat{\pars}} \max_{s \in \mathcal{S} \setminus \mathcal{S}_f} \bm{M}_s \cdot \hat{\pars}  \,, \quad \text{with } \nonumber \\
     \bm{M}_s \cdot \hat{\pars} = 0\,, \,\,\,\forall s \in \mathcal{S}_f \,,\quad \text{and } \|\hat{\pars} \| = 1 \,,
\end{align}
where $\mathcal{S}$ is the set of all possible structures, $\mathcal{S}_f$ is the designable set whose yield we wish to maximize, and where we constrain the length of the parameter vector via a linear `norm' $\| \pars \| = \sum \bm{\varepsilon} - \sum \bm{\mu}$ to maintain the linearity of the problem.
This optimization problem can be solved quickly and robustly with linear programming~\cite{boyd2004convex}.

If a solution $\hat{\pars}^\star_f$ to this problem exists with an optimal value
\begin{equation}
    x^\star_f = \max_{s \in \mathcal{S} \setminus \mathcal{S}_f} \left[\bm{M}_{s} \cdot \hat{\pars}^\star_f\right] < 0 \,,
\end{equation}
then $\mathcal{S}_f$ is asymptotically designable and $\hat{\pars}^\star_f$ is the optimal limiting direction ($\hat{\pars}^\star_f$ can then be renormalized with a quadratic norm, if desired).
On the other hand, if $x^\star = 0$, then there exist additional structures that cannot be suppressed while assembling $\mathcal{S}_f$, meaning that the set is not designable.
In practice, it is also often convenient to preprocess the optimization by setting any design variables that do not affect the target structures to $-\infty$, which corresponds physically to removing any particle species or bond type that is not required for assembly of the target(s).

Furthermore, it is often desirable to take into account additional constraints, which are not necessarily linear in the design parameters, for example, the total particle concentration
\begin{equation}
    \phi(\pars) = \sum_{s \in \mathcal{S}} n_s \rho_s(\pars) \,,
\end{equation}
where $n_s$ is the number of particles in structure $s$.
Fixing the total concentration at some desired maximal value $\phi_\star$ is a non-linear constraint that goes beyond the linear theory discussed in the main text and above.

To take fixed particle concentration into account during parameter optimization, we can first write the yield of a structure $s$ as
\begin{equation}
    Y_s(\pars) = \frac{\Omega_s e^{\M\cdot \pars}}{\sum_{s'} \Omega_{s'} e^{\bm{M}_{s'}\cdot \pars}} = \left[1 + R_s(\pars)\right]^{-1}\,,
\end{equation}
with 
\begin{equation}
    R_s(\pars) = \sum_{s' \neq s} \frac{\Omega_{s'}}{\Omega_s} e^{(\bm{M}_{s'}-\M)\cdot\pars} \,.
\end{equation}
Maximizing $Y_s(\pars)$ is clearly equivalent to minimizing $R_s(\pars)$.
Importantly, both $R_s(\pars)$ and $\phi(\pars)$ are convex functions.
This means that we can employ convex optimization~\cite{boyd2004convex, convexjl} to optimize structure yields at particle concentration $\phi \leq \phi_\star$.
Examples of yield optimizations under concentration constraints are shown below in Fig.~\ref{sfig:squaretraj}.

\subsection{Intersections of designable sets}
As mentioned in Methods, the correspondence between designable sets of structures and polyhedral faces implies that the intersection of two designable sets is also designable.
To see this, consider two designable sets $\mathcal{S}_f$ and $\mathcal{S}_g$, their corresponding faces $f$ and $g$, and their conic combination $f + g = \{\alpha x + \beta y \mid x\in f, y\in g\,, \alpha, \beta \geq 0\} \equiv  \{x + y \mid x\in f, y\in g\,\}$.
We will show that the designable set corresponding to $f+g$ is given by the intersection of $\mathcal{S}_f$ and $\mathcal{S}_g$.

First note that the designable set $\mathcal{S}_f$ consists of the structures $s\in \mathcal{S}$ for which $\M \cdot x = 0$ for all $x\in f$, or, more formally,
\begin{equation}
    \mathcal{S}_f = \{s\in\mathcal{S} \mid \text{$\M \cdot x = 0$ for all $x \in f$}\}\,.
\end{equation}
Similarly, $\mathcal{S}_{f+g}$ is given by
\begin{align}
    \mathcal{S}_{f + g} = \{s\in\mathcal{S} \mid \,&\text{$\M \cdot (x + y) = 0$} \nonumber \\
    &\text{for all $x \in f, y \in g$}\}\,.
\end{align}
Since $f$ and $g$ are faces of the cone, $\M \cdot x \leq 0$ and $\M \cdot y \leq 0$, which implies that $\M \cdot (x + y) = 0$ if and only if $\M \cdot x = 0$ and $\M \cdot y = 0$.
This means that a structure $s$ is in $\mathcal{S}_{f+g}$ if and only if $s\in\mathcal{S}_f$ and $s\in\mathcal{S}_g$, and therefore
\begin{equation}
    \mathcal{S}_{f + g} = \mathcal{S}_f \cap \mathcal{S}_g  \,. 
\end{equation}

Note that $f + g$ is not necessarily a face of the cone, but may only be a subset of a face.
If this is the case, then there exists at least one other face $h$, such that $\mathcal{S}_f \cap \mathcal{S}_g \cap \mathcal{S}_h = \mathcal{S}_f \cap \mathcal{S}_g$ and the complete face corresponding to $\mathcal{S}_f \cap \mathcal{S}_g$ is given by $f + g + h$.

The fact that the intersection of designable sets is again designable also has important implications for non-designable sets.
While thermodynamic constraints prevent high yield for any non-designable set of structures, the intersection property of designable sets guarantees that any non-designable set is contained in a unique \emph{minimal designable set}.
This is readily proved by noting that if there were two different minimal designable sets that contain the non-designable target(s), we could form their intersection to obtain a smaller containing set.

\subsection{Note on negative binding energies}
Fig.~3a of the main text shows that the allowed limit directions include directions where the binding energies approach $-\infty$.
In our convention, a negative binding energy for a certain bond means that this `bond' is actually repulsive and that structures that contain this bond are less likely to form.
In the limit $\varepsilon\to -\infty$, this means that a structure containing a `repulsive bond' cannot form and can be safely neglected.

The same approximation is already made implicitly in the enumeration step: we assume that only particles with complementary binding sites are able to bind -- assuming that the yield of all structures with undesired bonds is exactly zero corresponds to setting the binding energy of all those undesired bonds to $-\infty$.
Taking a "negative energy" limit is therefore equivalent to removing a bond from the binding rules.
If one wishes to enumerate only those designable structures that contain all bond types, one can simply add additional inequality constraints that force energies to be non-negative, i.e., a constraint of the form $\bm{\varepsilon} \geq 0$.

\subsection{Supplemental data on the example systems}
In this supplemental section, we briefly describe the architectures and some designable structures of additional example systems.

\begin{figure}
    \centering
    \includegraphics[width=1\textwidth]{figS1.pdf}
    \caption{Hasse diagram of the system shown in Fig.~1c of the main text, with the additional constraint that all four binding energies are equal. The twelve individually designable structures are shown next to their corresponding nodes in the diagram.}
    \label{sfig:smallcluster}
\end{figure}
\paragraph{Small clusters}
For completeness, we briefly discuss the designable structures resulting from the binding rules of the system in Fig.~1d of the main text.
This system only leads to 16 possible structures, but still has eight degrees of freedom to tune yields.
This large amount of control leads to all 16 structures being asymptotically designable. Note that we distinguish between a ``shape'' and a ``structure'': any two dimers, for example, are the same shape, but if the particle types differ at all, then they are different structures. Experimentally, it is much easier to differentiate between shapes than between structures, which is why in Fig.~1d of the main text we show the yield of the 7 distinct shapes.

However, the situation becomes more interesting when we introduce further constraints on the parameters, for example, by imposing that the four distinct programmable bonds have equal binding energies, which is a practical constraint often encountered in experimental systems.
We can easily incorporate this constraint simply by stopping to distinguish the different types of bonds that the structures contain.

In this case, there are twelve designable structures and 33 designable sets in the next level of the architecture, most of which correspond to pairs of structures.
The Hasse diagram, together with the individually designable structures, is shown in Fig.~\ref{sfig:smallcluster}.

\paragraph{Reconfigurable squares}
In the main text, we have shown that the binding rules in Fig.~3c lead to 677 possible structures, out of which seven non-monomer structures are designable if binding energies are uniform.
This means that it is possible to maximize the equilibrium yields of any of the seven structures simply by changing particle concentrations.
As mentioned in the main text, in the limit of high binding energy, one simply has to choose particle concentrations to be proportional to the stoichiometry of the desired target, assuming bonds are perfectly specific and crosstalk can be neglected~\cite{Murugan.2015}.

Here we show an explicit example of how to continuously toggle between the different structures.
With the help of convex optimization, as described above, we maximize the yields of every target structure while imposing a binding energy $\varepsilon = 18\,\mathrm{k_BT}$ and total particle concentration $\phi \leq 0.01 v_\mathrm{eff}$, where $v_\mathrm{eff}$ is the effective volume a bound particle can explore.
For simplicity, we assume the entropic partition function of the structures to be given by
\begin{equation}
    \Omega_s = \frac{2\pi v_\mathrm{eff}^{n_s - 1}}{\lambda^{D n_s} \sigma_s} \,.
\end{equation}
We then perform seven separate convex optimizations to find parameters that maximize the yield of each target.

\begin{figure}
    \centering
    \includegraphics[width=1\textwidth]{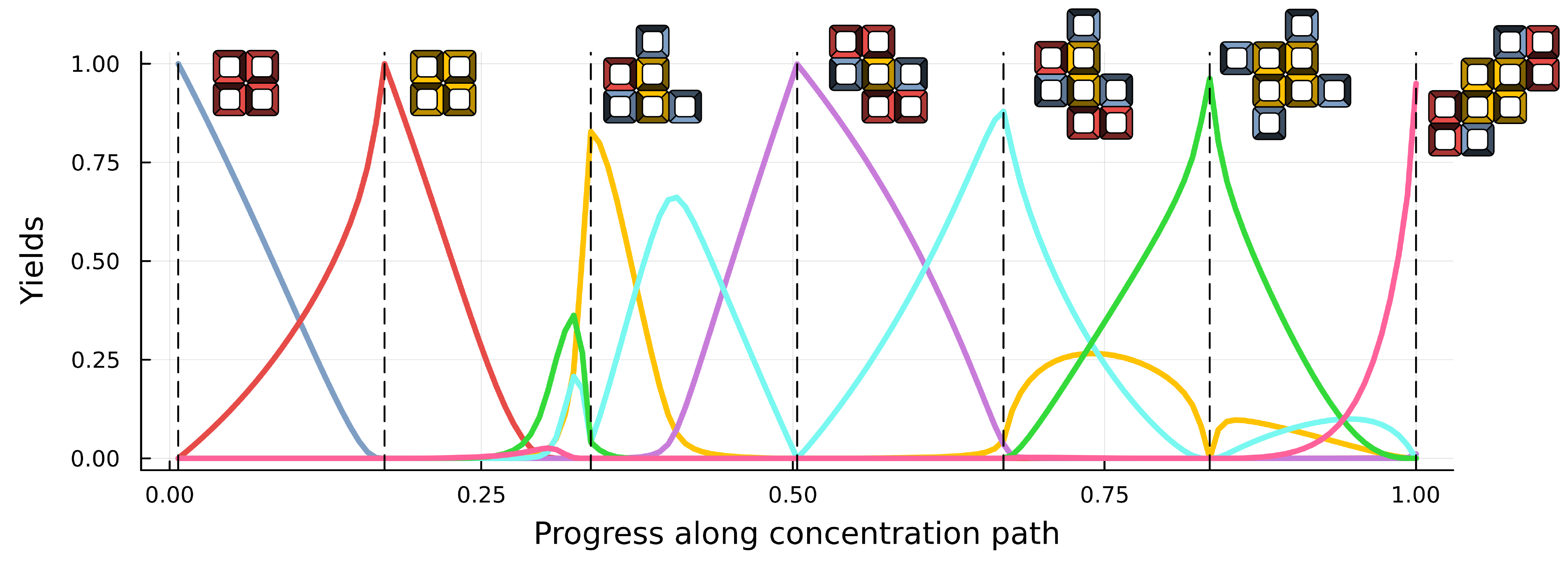}
    \caption{Reconfigurable assembly by tuning particle concentrations. Following a path in concentration space, obtained by linearly interpolating between the concentrations that maximize each target, we can toggle between the assembly of different designable structures. Vertical lines show the locations along the path that maximize the yields of the shown structures. The particle concentrations at these points are listed in Table~\ref{table:squretraj}. Optimizations were performed at binding energy $\varepsilon = 18\,\mathrm{kT}$ and total particle concentration $\phi \leq 0.01\, v_\mathrm{eff}$.}
    \label{sfig:squaretraj}
\end{figure}

After this, we compute the particle concentrations at each of the optimized parameters, which, as expected, correspond to the stoichiometry of each target (see Table~\ref{table:squretraj}).
We then create a path through the three-dimensional concentration space by linearly interpolating between the seven optimized concentrations.
Following this path will, one after the other, maximize the yield of every target.
This is shown in Fig.~\ref{sfig:squaretraj}, confirming that it is possible to ``toggle between'' different designable structures simply by tuning the relative concentrations of the particle species.
Note that the heights of the yield maxima depend on the imposed constraints, but are guaranteed to go to one in the limit of high binding energy, since all structures are individually designable.

\begin{table}
\begin{tabular}{c | c c c} 
     Target & $\phi_\mathrm{blue} / \phi$ & $\phi_\mathrm{red}/ \phi$ & $\phi_\mathrm{yellow}/ \phi$ \\
     \hline
     Red square & 0 & 0.999 & 0 \\ 
     \hline
     Yellow square & 0 & 0 & 0.999 \\
     \hline
     6 part. str. &  0.500 & 0.167 & 0.333 \\
     \hline
     7 part.  str. &  0.282 & 0.564 & 0.141 \\
     \hline
     Asym. 8 part.  str. &  0.371 & 0.371 & 0.248 \\ 
     \hline
     Sym. 8 part.  str.& 0.5 & 0 & 0.5 \\ 
     \hline
     10 part.  str.& 0.199 & 0.398 & 0.398 \\ 
\end{tabular}
\caption{Optimized relative concentrations $\phi_\alpha / \phi$ that maximize the yields of each designable structure shown in Fig.~3c and Fig.~\ref{sfig:squaretraj} at $\varepsilon=18 \, \mathrm{kT}$.}
\label{table:squretraj}
\end{table}

\paragraph{Coexisting crystals}
To construct an approximate constraint cone for the system in Fig.~3d of the main text, we generate all finite size clusters and periodic unit cells on small grids of size $i \times j$, where $i, j = 1...3$, which results in 1558 different grid configurations.
Computing the non-redundant constraints reveals that only 77 of these configurations correspond to non-redundant constraint planes. 

We then use linear programming (see above) to compute parameters aligned with the checkerboard unit cell (Fig.~3d(ii)), the tiling with holes (Fig.~3d(iii)), and the coexistence of the two (Fig.~3d(iv)).
By definition, choosing parameters strictly along a polyhedral face means that the resulting bulk phase has a free energy of zero.
To drive the assembly, we thus add a small supersaturation term, which in our language means adding a contribution $\pars_\perp$ normal to the constraint plane.
We choose $\pars_\perp$ to be the minimal-norm vector satisfying $\bm{M}_\mathrm{target} \cdot \pars_\perp = 0.1$.
The parameters used in simulation can thus be decomposed as
\begin{equation}
    \pars = \lambda \hat{\pars}+ \pars_\perp \,,
\end{equation}
where $\hat{\pars}$ is the limiting direction obtained through linear programming.

\begin{figure}
    \centering
    \includegraphics[width=1\textwidth]{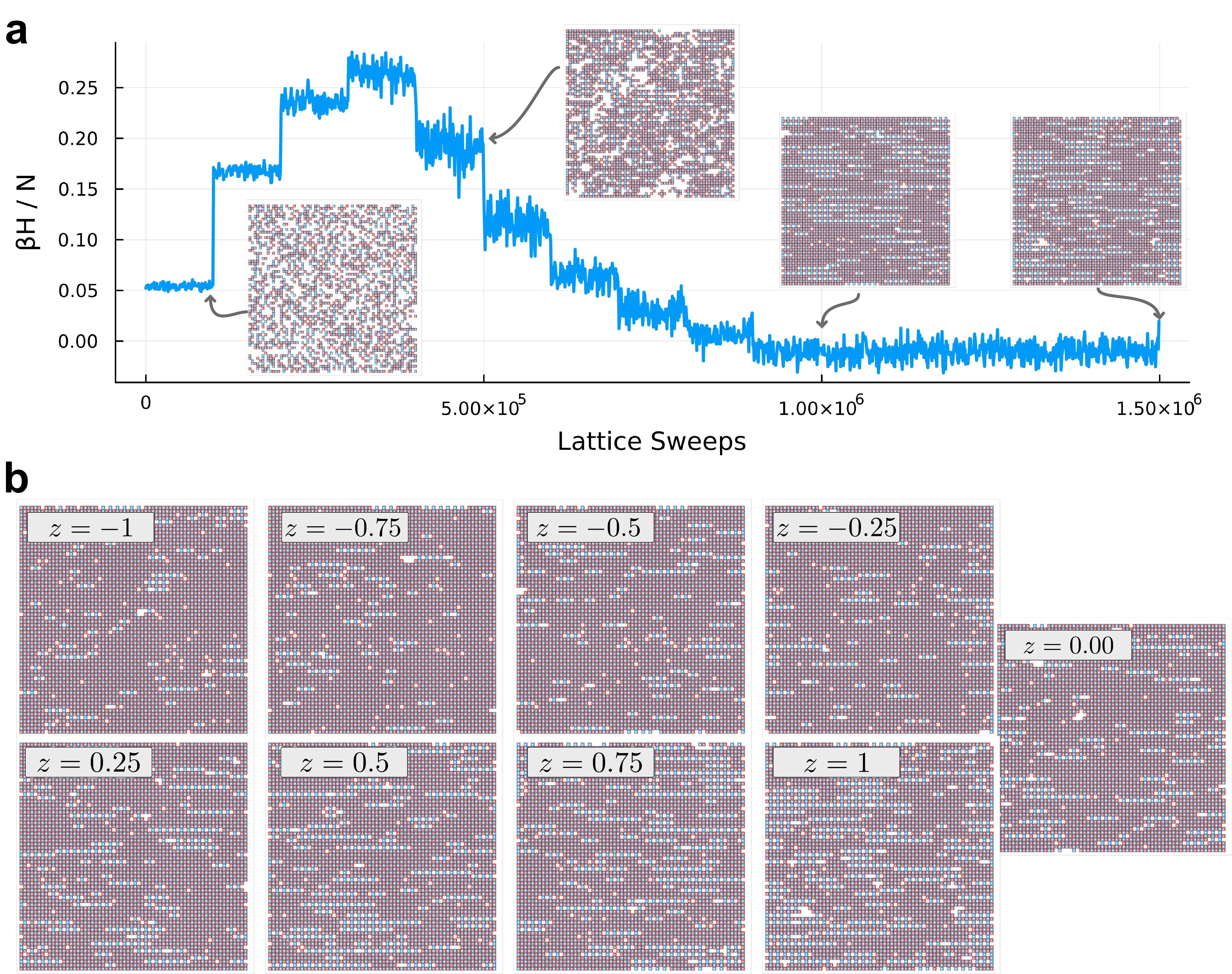}
    \caption{Supplementary information for crystallization simulations. (a) Lattice energy per site $\beta H / N$ as a function of time, measured in lattice sweeps and including system snapshots at different times. (b) Different coexisting ratios of the checkerboard and hole-tiling phases depending on the scale factor $z$. A section of the system snapshot at $z=1$ was used in the main text.}
    \label{sfig:lattice}
\end{figure}

We perform Monte Carlo simulations (see below) with three sets of parameters, leading to the assembly of the different phases. 
All used parameters are listed in Table~\ref{table:crystal_pars}.
All simulations were run for $1.5 \times 10^6$ lattice sweeps, where one lattice sweep attempts a Monte Carlo move at every site of the lattice.
To speed up equilibration, we gradually scale up $\lambda$ every 100,000 sweeps, starting from $\lambda=1$ and ending at $\lambda=15$.
We then keep simulating at $\lambda=15$ for 500,000 additional sweeps to ensure the configuration is stable.
A typical trajectory is shown in Fig.~\ref{sfig:lattice}a.

\begin{table}
\begin{tabular}{r | c c | c c c c c c c c} 
      & $\mu_\mathrm{blue}$ & $\mu_\mathrm{red}$ & $\varepsilon_{\mathrm{b,b}}^\mathrm{l-r}$ & $\varepsilon_{\mathrm{b,b}}^\mathrm{t-b}$ & $\varepsilon_{\mathrm{r,r}}^\mathrm{l-r}$ & $\varepsilon_{\mathrm{r,r}}^\mathrm{t-b}$ & $\varepsilon_{\mathrm{b,r}}^\mathrm{r-l}$ & $\varepsilon_{\mathrm{b,r}}^\mathrm{l-r}$ & $\varepsilon_{\mathrm{b,r}}^\mathrm{t-b}$ & $\varepsilon_{\mathrm{b,r}}^\mathrm{b-t}$\\
     \hline
     $\hat{\pars}_\mathrm{checker}$ & -0.577 & -0.577 & -100 & -100 & -100 & -100 &  0.289 & 0.289 & 0.289 & 0.289\\ 
     \hline
     $\hat{\pars}_\mathrm{holes}$ & -0.697 & -0.232 & 0.465 & -100 & -100 & -100 & -100 & 0.349 & -100 & 0.349 \\
     \hline
     $\hat{\pars}_\mathrm{coex}$ &  -0.688 & -0.229 & 0.459 & -100 & -100 & -100 & 0.115 & 0.344 & 0.115 & 0.344 \\
     \hline
     ${\pars_\perp}_\mathrm{checker}$ & 0.017 & 0.017 & 0 & 0 & 0 & 0 & 0.017 & 0.017 & 0.017 & 0.017\\ 
     \hline
     ${\pars_\perp}_\mathrm{holes}$ & 0.018 & 0.009 & 0.018 & 0 & 0 & 0 & 0 & 0.009 & 0 & 0.009 \\
     \hline
     ${\pars_\perp}_\mathrm{coex}$ & 0.047 & 0.041 & 0.012 & 0.00 & 0.00 & 0.00  & 0.035 & 0.041 & 0.035 & 0.041\\
     \hline
      ${\bm{\zeta}_\perp}_\mathrm{coex}$ &  0.045 & 0.006 & 0.077 & 0.00 & 0.00 & 0.00 & -0.032 & 0.006  & -0.032 & 0.006 \\
\end{tabular}
\caption{Chemical potentials $\mu$ and binding energies $\varepsilon$ (both in units of $k_\mathrm{B}T$) used for the Monte Carlo simulations of crystal formation. The subscripts $\mathrm{b,b}$,  $\mathrm{r,r}$, and $\mathrm{b, r}$ stand for interactions between blue-blue, red-red, and blue-red particles respectively, the superscripts $\mathrm{l-r}$, $\mathrm{r-l}$, $\mathrm{t-b}$, and $\mathrm{b-t}$ stand for interactions between the left and right, right and left, top and bottom, and bottom and top sides of the two particle species, respectively. Binding energies of bonds that are not present in the target structures are set to $-100$, effectively preventing the formation of the corresponding bonds.}
\label{table:crystal_pars}
\end{table}

For the coexistence of checkerboard and hole tiling, it is possible to tune the relative concentrations by moving normal to the associated constraint plane.
However, due to defects and non-trivial boundary interactions and anisotropic line tension between the two phases on the finite simulation lattice, achieving a desired relative concentration between the phases is not as straightforward as in the case for finite-size clusters, which do not interact with each other.
To tune the coexistence ratio, we add another contribution $\bm{\zeta}_\mathrm{\perp}$ to the parameters, again normal to the corresponding face, but designed to raise the concentration of one phase over the other.
This contribution was again obtained through simple linear algebra, as the least-norm vector that satisfies
\begin{equation}
    \left(\begin{array}{l}
    \bm{M}_{\rm checker}^T \\
    \bm{M}_{\rm hole}^T 
    \end{array}\right) \bm{\zeta}_\mathrm{\perp} = 
    \left(\begin{array}{c}
    0 \\
    1
    \end{array}\right) \,,
\end{equation}
and the parameters used for the coexistence can be written as $\pars = \lambda \hat{\pars}+ \pars_\perp + z \bm{\zeta}_\perp$.
Figure~\ref{sfig:lattice}b shows system snapshots as $\bm{\zeta}_\mathrm{\perp}$ is scaled by a factor $z = -1$ to $z = 1$.
The system snapshot obtained at  $z = 1$ is the one shown in the main text.

\begin{figure}
    \centering
    \includegraphics[width=1\textwidth]{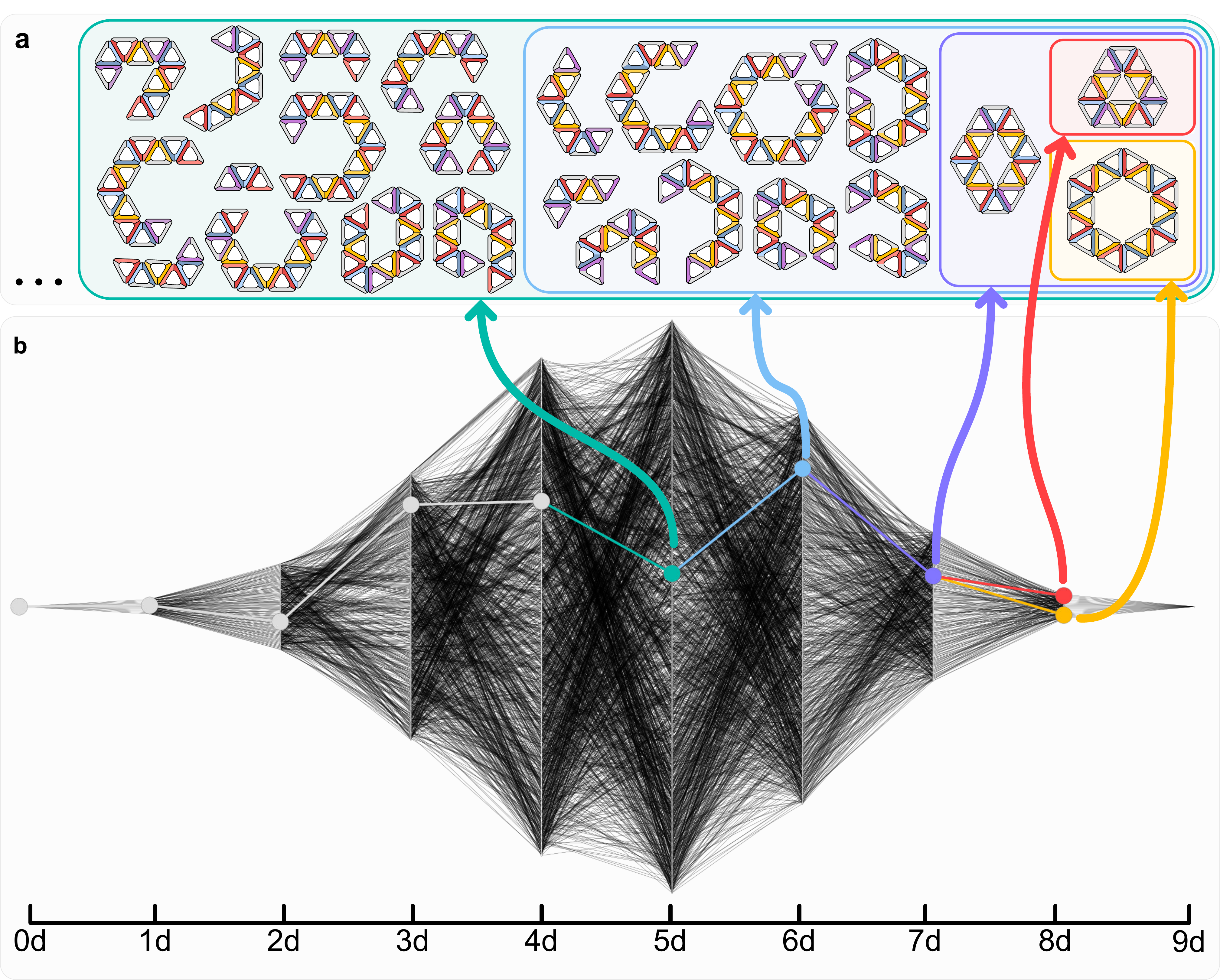}
    \caption{Designable sets and Hasse diagram of the three ring system. (a) Some of the structures that are allowed by the binding rules in Fig.~1b of the main text. Structures are grouped into hierarchical designable sets, showing the containment relations. (c) The Hasse diagram corresponding to the nine-dimensional constraint cone. The nodes corresponding to the designable sets in (a) are highlighted in color.}
    \label{sfig:rings}
\end{figure}

\paragraph{Three rings}
Here we show the full Hasse diagram for the binding rules shown in Fig.~1b of the main text, leading to the assembly of the hexagonal, rhomboid, and triangular rings.
Figure~\ref{sfig:rings}a shows some of the designable sets that follow from the binding rules, highlighting the fact that the designable sets are nested within each other.
These sets correspond to the highlighted nodes of the Hasse diagram shown in Fig.~\ref{sfig:rings}b.
The full diagram consists of 1672 designable sets and was generated in less than a second using the algorithm described in Methods.

\subsection{Monte Carlo simulations}
The Monte Carlo simulations were performed on a square lattice of size $64\times 64$.
Following the notation of Ref.~\cite{Murugan.2015kjd}, we denote a specific lattice configuration by $\sigma_i^\alpha$, where $\sigma_i^\alpha = 1$ if lattice site $i$ is occupied by species $\alpha$ and $\sigma_i^\alpha = 0$ otherwise.
$\alpha = 0, 1, 2$ indexes the particle species and $\alpha = 0$ corresponds to empty sites. 
The energy of a specific lattice configuration can then be written as
\begin{equation}
    -\beta H(\sigma^\alpha_i) = \sum_{\langle i, j\rangle}\sum_{\alpha, \beta}\varepsilon_{\alpha \beta, \delta(i,j)} \sigma_i^\alpha \sigma_j^\beta + \sum_i\sum_\alpha \mu_\alpha \sigma_i^\alpha \,,
\end{equation}
where $\varepsilon_{\alpha\beta, \delta(i, j)}$ denotes the binding energy between species $\alpha$ and $\beta$.
Since binding energies between different sides of the particles are different, this also depends on the directionality of the contact, which is labeled by $\delta(i, j)$, which is given by the positions of the lattice sites $i$ and $j$ the contacting particles are located at.
$\mu_\alpha$ is the vector of chemical potentials of the particle species.
In the main text, $\mu_\alpha$ and the nonzero elements of $\varepsilon_{\alpha\beta, \delta}$ are always referred to together as the ``design parameters'' $\bm{\xi}$.
For consistency with the main text, we also define attractive interactions as having a positive binding energy.

We simulate this lattice model using the Metropolis algorithm.
Every Monte Carlo move consists of choosing a lattice site and randomly switching out the particle species at this site.
A move is accepted with probability $\min(1, e^{-\beta \Delta H})$ where $\Delta H$ is the energy difference caused by the proposed move.

\subsection{Supplemental experimental data}
\begin{figure*}[!bh]
    \centering
    \includegraphics[width=\linewidth]{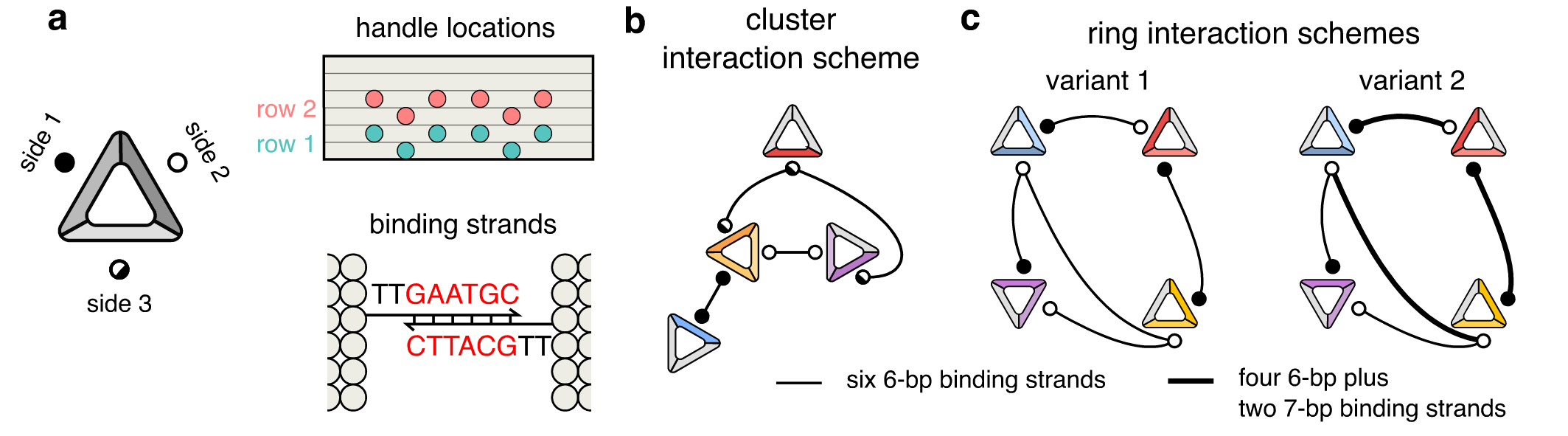}
\caption{\textbf{Design of self-assembly experiments}. (a) Sketch of where interactions are located on the triangular monomer. Each side of the triangle has a set of 12 staples that can be extruded from the face of the DNA-origami particle. (b) Binding scheme used for the cluster experiments shown in Fig. 1c of the main text. Sequences for these interactions are shown in Table~\ref{tab:clusters-ints} (c) Binding schemes for the ring experiments shown in Fig. 4 of the main text. For these experiments we have two variants, the second of which we intentionally make the interactions with the purple particle weaker relative to the other interactions. Sequences for these interactions are shown in Table~\ref{tab:sideinteractions} and \ref{tab:sideinteractions-strong}.}
    \label{Fig:sfig-dimercoord}
\end{figure*}

\begin{center}
\begin{figure*}[!h]
\centering
\includegraphics[width=\linewidth]{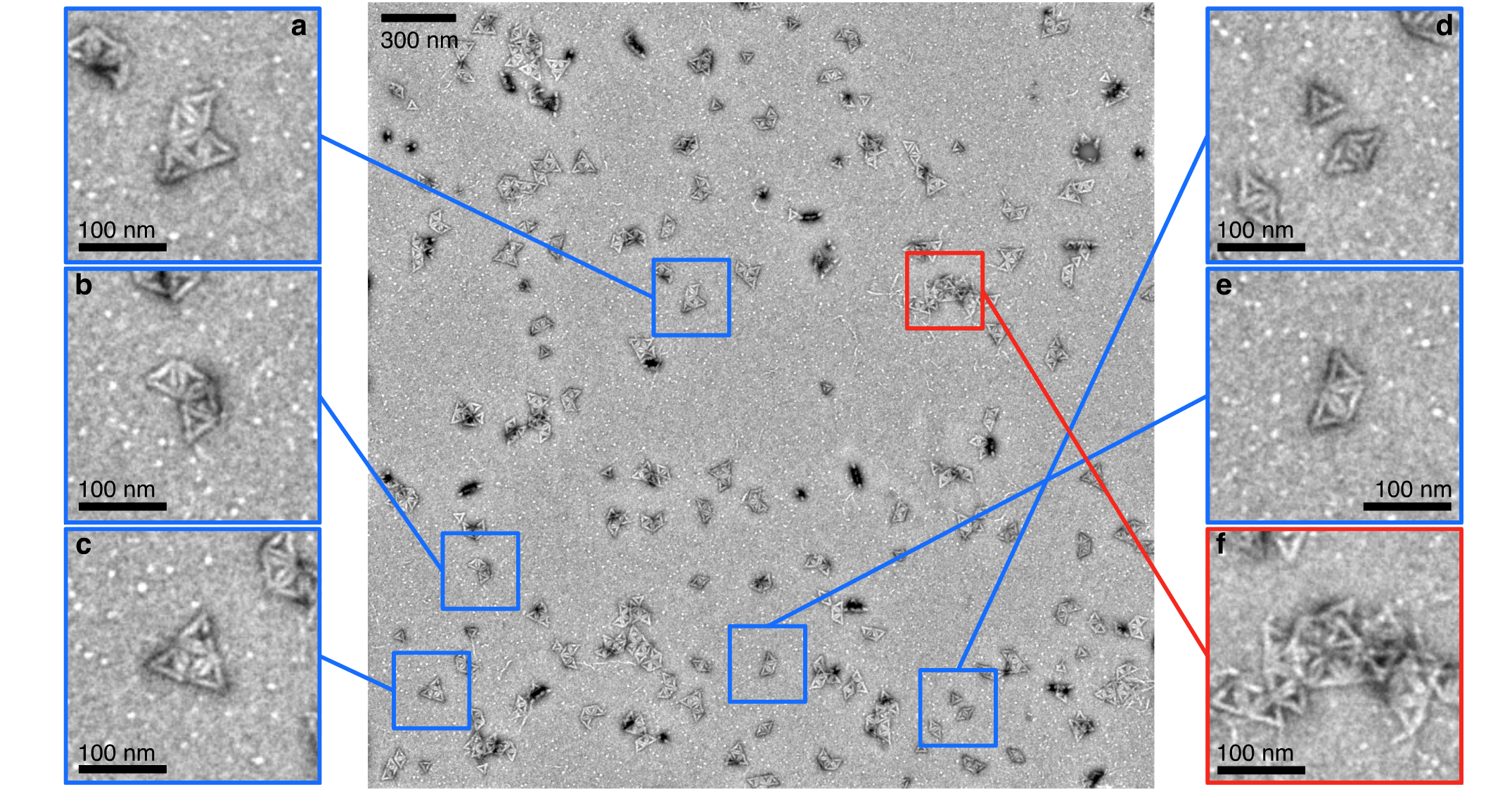}
\caption{\textbf{Representative TEM image from a cluster experiment.} The center image is a 1500x1500 pixel subregion from a 2560x2048 pixel experimental TEM image. Panels (a)-(e) show different clusters that can be seen in the image, the only one that is missing here is the linear tetramer shown in Fig.~1c. Occasionally we find overlapping clusters, shown in (f). These most likely do to sample preparation and are not included in our statistics. The TEM image has been bandpass filtered to enhance contrast.}
\label{sfig:cluster-image}
\end{figure*}    
\end{center}

\begin{center}
\begin{figure*}[!h]
\centering
\includegraphics[width=\linewidth]{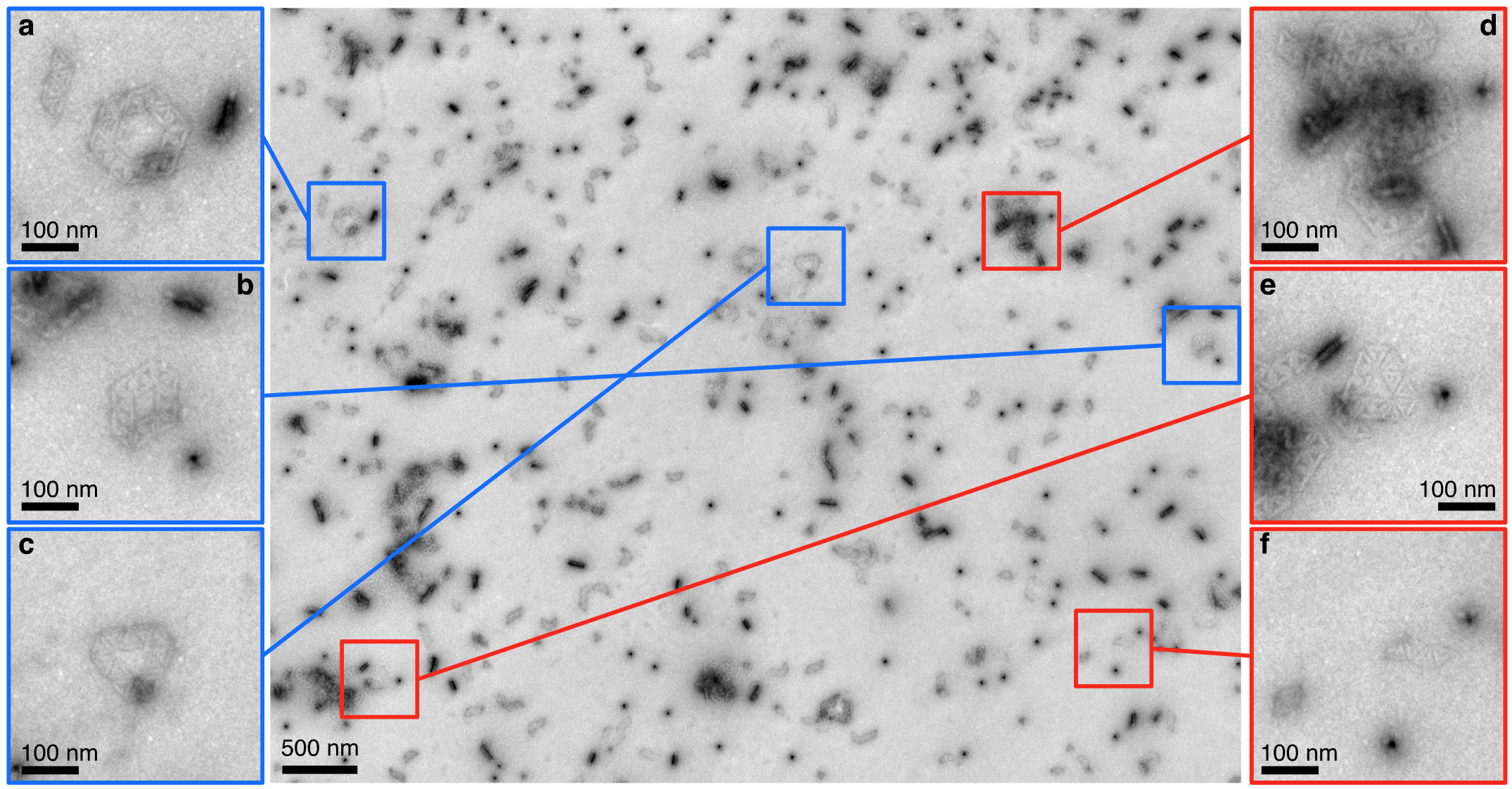}
\caption{\textbf{Representative TEM image from a ring experiment.} The center image is the size of the field of view that we take experimental TEM images. In this image, we can see several different features of our assembly experiments. Panels (a)-(c) show the three varieties of ring structures that we targeted. However, the majority of the field of view is filled with incomplete assemblies or artifacts from sample preparation. For instance, panel (d) shows an agglomeration that we typically ascribe to the clumping of structures induced by staining the sample with UFo and then drying it. We also observe structures that overshoot an intended target, as in panel (e), due to the flexibility of the DNA-origami intersubunit bonds; assemblies of triangles are not necessarily constrained to be planar. Finally, the predominant things we find are monomers or small oligomers, shown in panel (f).}
\label{sfig:ring-image}
\end{figure*}    
\end{center}

Here we provide supplemental experimental data for the experimental results shown in Fig.~1c and Fig.~4b of the main text.
Figure~\ref{Fig:sfig-dimercoord} contains details about the design of the building blocks and the programmable bonds.
Table~\ref{tab:cluster-conditions} contains the experimental conditions and structure counts for the small cluster experiments shown in Fig.~1c.
Table~\ref{tab:rings-conditions} contains the experimental conditions and structure counts for the ring experiments shown in Fig.~4b.
Figure~\ref{sfig:cluster-image} and Fig.~\ref{sfig:ring-image} contain representative TEM images from the small cluster and ring experiments respectively.
Table~\ref{tab:clusters-ints}, Table~\ref{tab:sideinteractions}, and Table~\ref{tab:sideinteractions-strong} show the interaction sequences for the different experiments.

\begin{longtable}{p{0.08\linewidth}  p{0.08\linewidth} p{0.08\linewidth} p{0.08\linewidth} p{0.08\linewidth} p{0.08\linewidth} p{0.08\linewidth} p{0.08\linewidth} p{0.08\linewidth} p{0.08\linewidth} }
\caption{\textbf{Experimental conditions for cluster assemblies.} $c_{\rm 0}$ is the concentration for the blue, yellow, and purple particles. $c_{\rm red}$ is the concentration for the red particle. $N_{\rm mono}$ is the counts of monomers, Fig.~\ref{sfig:cluster-image}d. $N_{\rm dimer}$ is the counts of dimers, Fig.~\ref{sfig:cluster-image}d. $N_{\rm trimer}$ is the counts of trimers, Fig.~\ref{sfig:cluster-image}e. $N_{\rm triangle}$ is the counts of triangular tetramers, Fig.~\ref{sfig:cluster-image}c. $N_{\rm concave}$ is the counts of concave tetramers, Fig.~\ref{sfig:cluster-image}b. $N_{\rm mono}$ is the counts of linear tetramers, Fig.~1c. $N_{\rm boat}$ is the counts of fully formed pentamers, Fig.~\ref{sfig:cluster-image}a.
}
\label{tab:cluster-conditions} \\
$c_0$ [nM]	&	$c_{\rm red}$ [nM]	&	MgCl$_2$ [mM]	&	$N_{\rm mono}$	&	$N_{\rm dimer}$	&	$N_{\rm trimer}$	&	$N_{\rm triangle}$	&	$N_{\rm concave}$	&	$N_{\rm line}$	&	$N_{\rm boat}$		\\
\hline
\endfirsthead
$c_0$ [nM]	&	$c_{\rm red}$ [nM]	&	MgCl$_2$ [mM]	&	$N_{\rm mono}$	&	$N_{\rm dimer}$	&	$N_{\rm trimer}$	&	$N_{\rm triangle}$	&	$N_{\rm concave}$	&	$N_{\rm line}$	&	$N_{\rm boat}$		\\
\hline
\endhead
2	&	0	&	20	&	159	&	131	&	339	&	2	&	5	&	0	&	0	\\
2	&	1	&	20	&	157	&	94	&	246	&	72	&	75	&	7	&	13	\\
2	&	2	&	20	&	172	&	97	&	171	&	125	&	114	&	17	&	72	\\
2	&	3	&	20	&	201	&	104	&	164	&	69	&	74	&	28	&	122	\\
2	&	4	&	20	&	75	&	30	&	49	&	30	&	20	&	15	&	90	\\
2	&	5	&	20	&	189	&	49	&	39	&	25	&	18	&	17	&	141	

\end{longtable}

\begin{longtable}{p{0.12\linewidth} | p{0.14\linewidth} p{0.09\linewidth} p{0.16\linewidth}p{0.20\linewidth} p{0.05\linewidth} p{0.05\linewidth} p{0.05\linewidth} }
\caption{\textbf{Experimental conditions for ring assemblies.} $c_{\rm R,B,Y}$ is the concentration for the red, blue, and yellow particles. $c_{\rm P}$ is the concentration for the purple particle. Under ``Interaction type'' we use ``Equal'' to denote the six-base interactions in Table~\ref{tab:sideinteractions} and ``Biased'' to denote the seven-base biased interactions in Table~\ref{tab:sideinteractions-strong}. $N_t$, $N_r$, $N_h$ are the counts for the triangular, rhombic, and hexagonal rings that we observed.} \label{tab:rings-conditions} \\
Exp. No.	&	$c_{\rm R,B,Y}$ [nM]	&	$c_{\rm P}$ [nM]	&	MgCl$_2$ [mM]   &	Interaction type & $N_t$ & $N_r$ & $N_h$ \\
\hline
\endfirsthead
Exp. No.	&	$c_{\rm R,B,Y}$ [nM]	&	$c_{\rm P}$ [nM]	&	MgCl$_2$ [mM]   &	Interaction type & $N_t$ & $N_r$ & $N_h$ \\
\hline
\endhead
1	&	10	&	0	&	30	&	Equal	&	0	&	0	&	35	\\
2	&	10	&	2	&	30	&	Equal	&	7	&	42	&	58	\\
3	&	10	&	4	&	30	&	Equal	&	36	&	24	&	5	\\
4	&	10	&	6	&	30	&	Equal	&	56	&	35	&	3	\\
5	&	10	&	8	&	30	&	Equal	&	95	&	59	&	3	\\
6	&	10	&	10	&	30	&	Equal	&	65	&	32	&	2	\\
7	&	10	&	1	&	20	&	Equal	&	2	&	14	&	80	\\
8	&	10	&	2	&	20	&	Equal	&	35	&	67	&	67	\\
9	&	10	&	3	&	30	&	Equal	&	34	&	47	&	16	\\
10	&	10	&	5	&	25	&	Biased	&	1	&	6	&	48	\\
11	&	10	&	10	&	25	&	Biased	&	2	&	22	&	41	\\
12	&	10	&	15	&	25	&	Biased	&	7	&	17	&	35	\\
13	&	10	&	20	&	25	&	Biased	&	5	&	13	&	28	\\
14	&	10	&	10	&	25	&	Biased	&	31	&	65	&	83	\\
15	&	10	&	20	&	25	&	Biased	&	34	&	42	&	33	\\
16	&	5	&	20	&	25	&	Biased	&	24	&	50	&	22	\\
17	&	5	&	30	&	25	&	Biased	&	7	&	22	&	8	\\
18	&	10	&	10	&	30	&	Biased	&	21	&	46	&	49	\\
19	&	10	&	20	&	30	&	Biased	&	30	&	66	&	40	

\end{longtable}

\section{Yield estimate for ring assemblies}

We first estimate the absolute yield of our ring assemblies from our TEM micrographs. We look at images like those shown in Fig.~\ref{sfig:ring-image} and count how many closed ring structures there are as well as how many subunits are in the field of view. We looked in detail at experiments No.~14 and No.~17 (see Table~\ref{tab:rings-conditions}) to get a rough estimate for the yield, noting that qualitatively other experimental conditions looked similar to these. For both experiments, we took 24 micrographs from which we counted the total number of closed rings. Using the first micrograph from each experiment, we counted the total number of subunits, about 620 and 1,650 respectively, and used this to estimate the number of subunits in all the images, 14,880 and 39,600 respectively. From this, we can compute either the total number of subunits in closed rings compared to the total number of subunits or the total number of ring structures compared to the number of subunits. These two estimates give yields of 18\% and 1.2\% for experiment No.~14 and 1.3\% and 0.1\% for experiment No.~17.

\begin{center}
\begin{figure*}[!h]
\centering
\includegraphics[width=\linewidth]{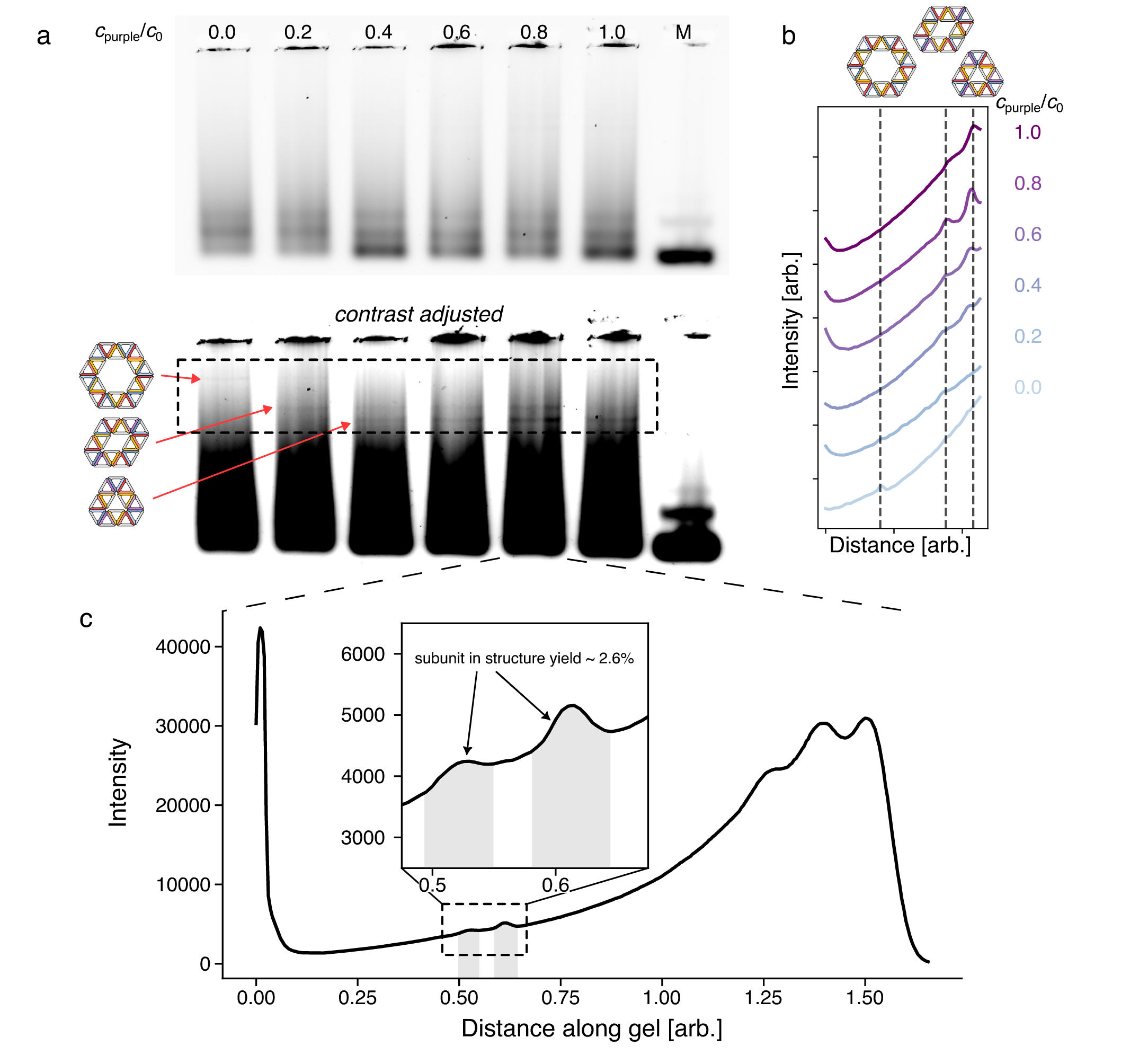}
\caption{\textbf{Estimating yield from an agarose gel.} (a) An agarose gel scan for assemblies with different ratios of purple particles to other particles. The lane label `M' is a monomer control lane. The lower gel is the same image, but contrast-enhanced to allow visualization of three faint bands corresponding to the different ring structures. (b) Plot of the intensity curves in the regions near the assembly bands for each assembly condition. Curves have been shifted up and down to allow easier comparison. The dotted lines are placed at the locations of the assembly peaks. (c) Plot of the intensity along the $c_{\mathrm{purple}} / c_0=0.8$ lane. The inset shows the curve near the triangle and rhombus ring location. The integrated intensity of the ring bands compared to the integrated intensity for the gel lane is 2.6\%.}
\label{sfig:ring-yield}
\end{figure*}    
\end{center}

For another estimate of the absolute yield, we perform agarose gel electrophoresis to separate different-sized structures from our assembly. In Fig.~\ref{sfig:ring-yield}a we show a gel scan from experiments No. 1 to No. 6. Only after adjusting the contrast do we see inklings of bands corresponding to the ring structures. Despite their low intensity, plots of the intensity for the different lanes can still show the presence of the three types of rings at different ratios of $c_\mathrm{purple}/c_0$ (Fig.~\ref{sfig:ring-yield}b). To estimate the yield of these assemblies, we consider the gel lane for $c_\mathrm{purple}/c_0=0.8$ and calculate the integrated intensity for the entire lane as well as the peaks for the rhombus and triangle rings; taking the ratio of the structure peak intensity to the lane intensity gives a yield of 2.6\% which is equivalent to the measure of the number of subunits compared to all subunits from before. To get a measure of the yield as the number of formed structures compared to all subunits, we account for the number of subunits in the rhombus and triangle, getting a yield of 0.2\%. These are comparable to our estimate from the TEM micrographs. Since our measures of relative yield are with respect to the structure count and not the subunit counted, we use the lower estimate for our yield.

\pagebreak

\begin{longtable}{p{0.08\linewidth} p{0.05\linewidth} p{0.05\linewidth} p{0.12\linewidth}p{0.12\linewidth} p{0.12\linewidth} p{0.12\linewidth} p{0.12\linewidth} p{0.12\linewidth}}
\caption{\textbf{Side interactions for small cluster assembly.} A list of the set of six interaction sequences that make up a side interaction of a monomer. The sequences are for self-complimentary side interactions, e.g. Position 1 binds to Position 6, Position 2 binds to Position 5, and Position 3 binds to Position 4.} \label{tab:clusters-ints} \\
Particle	&	Side	&	Row	&	Position 1	&	Position 2	&	Position 3	&	Position 4	&	Position 5	&	Position 6	\\
\hline
\endfirsthead
Particle	&	Side	&	Row	&	Position 1	&	Position 2	&	Position 3	&	Position 4	&	Position 5	&	Position 6	\\
\hline
\endhead
Blue	&	1	&	2	&	CAATAG	&	TGATTG	&	CTAGGA	&	CACATC	&	ACGAAG	&	ACCTGA	\\ 
Yellow	&	1	&	2	&	TCAGGT	&	CTTCGT	&	GATGTG	&	TCCTAG	&	CAATCA	&	CTATTG	\\
	&	2	&	2	&	ATGACA	&	TACAGG	&	AACCTA	&	GAGACA	&	GACAGA	&	ACTAAC	\\
	&	3	&	2	&	GTACAT	&	AGTCAG	&	CGATGG	&	CTTACT	&	AGTATC	&	GTATGT	\\
Purple	&	2	&	2	&	GTTAGT	&	TCTGTC	&	TGTCTC	&	TAGGTT	&	CCTGTA	&	TGTCAT	\\
	&	3	&	2	&	GTACAT	&	AGTCAG	&	CGATGG	&	CTTACT	&	AGTATC	&	GTATGT	\\
Red	&	3	&	2	&	ACATAC	&	GATACT	&	AGTAAG	&	CCATCG	&	CTGACT	&	ATGTAC	

\end{longtable}

\begin{longtable}{p{0.08\linewidth} p{0.05\linewidth} p{0.05\linewidth} p{0.12\linewidth}p{0.12\linewidth} p{0.12\linewidth} p{0.12\linewidth} p{0.12\linewidth} p{0.12\linewidth}}
\caption{\textbf{Side interactions for ring assembly: Variant 1.} A list of the set of six interaction sequences that make up a side interaction of a monomer. The sequences are for self-complimentary side interactions, e.g. Position 1 binds to Position 6, Position 2 binds to Position 5, and Position 3 binds to Position 4.} \label{tab:sideinteractions} \\
Particle	&	Side	&	Row	&	Position 1	&	Position 2	&	Position 3	&	Position 4	&	Position 5	&	Position 6	\\
\hline
\endfirsthead
Particle	&	Side	&	Row	&	Position 1	&	Position 2	&	Position 3	&	Position 4	&	Position 5	&	Position 6	\\
\hline
\endhead
Red	&	1	&	2	&	CAATAG	&	TGATTG	&	CTAGGA	&	CACATC	&	ACGAAG	&	ACCTGA	\\
	&	2	&	2	&	AGATAG	&	TTCCTG	&	TTCCAT	&	GATATG	&	ATGCAC	&	AACATT	\\
Blue	&	1	&	2	&	GTTAGT	&	TCTGTC	&	TGTCTC	&	TAGGTT	&	CCTGTA	&	TGTCAT	\\
	&	1	&	1	&	GTACGA	&	TCACAG	&	TGAGAA	&	GAATCT	&	GGAATA	&	AGATGC	\\
	&	2	&	2	&	AATGTT	&	GTGCAT	&	CATATC	&	ATGGAA	&	CAGGAA	&	CTATCT	\\
Yellow	&	1	&	2	&	ATCTAC	&	CTCAAG	&	CTGAAT	&	CAGAAT	&	TAATCG	&	GGAACT	\\
	&	1	&	1	&	GCATCT	&	TATTCC	&	AGATTC	&	TTCTCA	&	CTGTGA	&	TCGTAC	\\
	&	2	&	2	&	AATGTT	&	GTGCAT	&	CATATC	&	ATGGAA	&	CAGGAA	&	CTATCT	\\
Purple	&	1	&	2	&	AGTTCC	&	CGATTA	&	ATTCTG	&	ATTCAG	&	CTTGAG	&	GTAGAT	\\
	&	2	&	2	&	ATGACA	&	TACAGG	&	AACCTA	&	GAGACA	&	GACAGA	&	ACTAAC	
\end{longtable}

\begin{longtable}{p{0.08\linewidth} p{0.05\linewidth} p{0.05\linewidth} p{0.12\linewidth}p{0.12\linewidth} p{0.12\linewidth} p{0.12\linewidth} p{0.12\linewidth} p{0.12\linewidth}}
\caption{\textbf{Side interactions for ring assembly: Variant 2.} A list of the set of six interaction sequences that make up a side interaction of a monomer. The sequences are for self-complimentary side interactions, e.g. Position 1 binds to Position 6, Position 2 binds to Position 5, and Position 3 binds to Position 4.} \label{tab:sideinteractions-strong} \\
Particle	&	Side	&	Row	&	Position 1	&	Position 2	&	Position 3	&	Position 4	&	Position 5	&	Position 6	\\
\hline
\endfirsthead
Particle	&	Side	&	Row	&	Position 1	&	Position 2	&	Position 3	&	Position 4	&	Position 5	&	Position 6	\\
\hline
\endhead
Red	&	1	&	2	&	CAATAG	&	TGATTG	&	CTAGGAG	&	CACATCG	&	ACGAAG	&	ACCTGA	\\
	&	2	&	2	&	GGATAA	&	TCATCC	&	AGATTCG	&	TTCTCAG	&	ACTGAG	&	AGAGAT	\\
Blue	&	1	&	2	&	AGTTCC	&	CGATTA	&	ATTCTGG	&	ATTCAGG	&	CTTGAG	&	GTAGAT	\\
	&	1	&	1	&	ATGACA	&	TACAGG	&	AACCTA	&	GAGACA	&	GACAGA	&	ACTAAC	\\
	&	2	&	2	&	TCAGGT	&	CTTCGT	&	CGATGTG	&	CTCCTAG	&	CAATCA	&	CTATTG	\\
Yellow	&	1	&	2	&	ATCTAC	&	CTCAAG	&	CCTGAAT	&	CCAGAAT	&	TAATCG	&	GGAACT	\\
	&	1	&	1	&	AGATAG	&	TTCCTG	&	TTCCAT	&	GATATG	&	ATGCAC	&	AACATT	\\
	&	2	&	2	&	ATCTCT	&	CTCAGT	&	CTGAGAA	&	CGAATCT	&	GGATGA	&	TTATCC	\\
Purple	&	1	&	1	&	AATGTT	&	GTGCAT	&	CATATC	&	ATGGAA	&	CAGGAA	&	CTATCT	\\
	&	2	&	1	&	GTTAGT	&	TCTGTC	&	TGTCTC	&	TAGGTT	&	CCTGTA	&	TGTCAT	

\end{longtable}